\documentclass[useAMS,usenatbib,referee,epsfig,epsf]{mn2e}
\usepackage{times,epsfig}

\title[First Kepler results on compact pulsators VIII]
{First \emph{Kepler} results on compact pulsators VIII: Mode identifications
via period spacings in $g-$mode pulsating Subdwarf B stars}

\author[M.D. Reed et al.]{
 M.~D.~Reed,$^1$\thanks{E-mail:MikeReed@missouristate.edu}
A.~Baran,$^{2,3}$ A.~C.~Quint,$^1$
S.~D.~Kawaler,$^2$ S.~J.~O'Toole,$^4$ J.~Telting,$^5$\newauthor
S.~Charpinet,$^6$
C.~Rodr\'iguez-L\'opez,$^{7,8}$, R.~H.~\O stensen,$^9$
J.~L.~Provencal,$^{10}$\newauthor E.~S.~Johnson,$^2$
S.~E.~Thompson,$^{11}$ C.~Allen,$^{12}$ C.~K.~Middour,$^{12}$
H.~Kjeldsen,$^{13}$ and\newauthor J.~Christensen-Dalsgaard$^{13}$\\
 $^1$Department of Physics, Astronomy and Materials Science,
 Missouri State University, 901 S. National, Springfield, MO 65897, USA \\
 $^2$Department of Physics and Astronomy, Iowa State University, Ames, IA 50011, USA\\
$^3$Suhora Observatory and Krakow Pedagogical University, ul. Podchor\c{a}\.{z}ych 2,30-084 Krak\'{o}w, Poland\\
$^4$Anglo-Australian Observatory, PO Box 296, Epping NSW 1710, Australia\\
$^5$Nordic Optical Telescope, 38700 Santa Cruz de La Palma, Spain\\
$^6$Laboratoire d'Astrophysique de Toulouse-Tarbes, Universit\'{e} de Toulouse, CNRS, 14 Av. E. Belin, 31400 Toulouse, France\\
$^7$Departemento de F\'isica Aplicada, Univ. de Vigo, Campus Lagoas-Marcosende s/n, 36310 Vigo, Spain\\
$^8$ Department of Physics and Astronomy, University of Delaware, 217 Sharp Lab, Newark DE 19716, USA\\
$^9$Instituut voor Sterrenkunde, Katholieke Universiteit Leuven, Celestijnenlaan 200 D, 3001 Leuven, Belgium\\
$^{10}$ Delaware Asteroseismic Research Center, Mt. Cuba Observatory, Greenville, DE, USA\\ Department of Physics and Astronomy, University of Delaware, Newark, DE, USA\\
$^{11}$ SETI Institute/NASA Ames Research Center, Moffett Field, CA 94035, USA\\
$^{12}$Orbital Sciences Corporation/NASA Ames Research Center, Moffett Field, CA 94035, USA\\
$^{13}$Department of Physics, and Astronomy, Building 1520, Aarhus University, 8000 Aarhus C, Denmark}

\date{Accepted
      Received }

\begin{document}

\maketitle

\begin{abstract}
We investigate the possibility of nearly-equally spaced periods in 13 hot 
subdwarf B (sdB) stars observed with the \emph{Kepler} spacecraft and one
observed with CoRoT. 
Asymptotic limits for gravity ($g-$)mode pulsations provide relationships between
equal period spacings of modes with differing degrees $\ell$ and relationships
between periods of the same radial order $n$ but differing degrees
$\ell$. Period transforms, Kolmogorov-Smirnov tests, and linear least-squares
fits have been used to
detect and determine the significance of equal period spacings. We have also
used  Monte Carlo simulations to estimate the likelihood that the detected
spacings could be produced randomly. 

Period transforms for nine of the \emph{Kepler} stars indicate $\ell =1$
period spacings, with five also showing peaks for $\ell =2$ modes. 12
stars indicate $\ell =1$ modes using the Kolmogorov-Smirnov test while
another shows solely $\ell =2$ modes. Monte Carlo results indicate that
equal period spacings are significant in 10 stars above 99\% confidence and 
13 of the 14 are above 94\% confidence. For 12 stars, the various methods
find consistent regular period spacing values to within the errors, two others
show some inconsistencies, likely caused by binarity, and the last has 
significant detections but the mode assignment disagrees between methods.

We use asymptotic period spacing relationships to associate observed periods
of variability with pulsation modes for $\ell =1$ and $2$. 
From the \emph{Kepler} first year survey sample of 13
multiperiodic $g-$mode pulsators, five stars have several consecutive overtones
making period spacings easy to detect, six others have fewer consecutive
overtones but period spacings are readily detected,
and two stars show marginal indications of
equal period spacings.
We also examine a $g-$mode sdB pulsator observed by CoRoT
with a rich pulsation spectrum and our tests detect regular period spacings.

We use Monte Carlo simulations to estimate the significance of the detections
in individual stars. From the simulations it is determined that regular period
spacings in 10 of the 14 stars is very unlikely to be random, another two
are moderately unlikely to be random and two are mostly unconstrained.

We find a common $\ell =1$ period spacing spanning a range from 231 to
272~s allowing us to correlate pulsation modes with 222 periodicities
and that the $\ell =2$ period spacings are related to the 
$\ell =1$ spacings by the asymptotic relationship $1/\sqrt{3}$.
We briefly discuss the impact of equal period spacings which indicate low-degree
modes with a lack of significant mode trappings.
\end{abstract}


\section{Introduction}
Asteroseismology is the process in which stellar pulsations are used to 
discern the physical condition of stars. The process includes matching
stellar models to observations, associating periodicities with
pulsation modes, and examining where the models succeed and fail. Slight
mismatches between observations and models can  provide insights to
new physics or add constraints to previously assumed conditions. Examples
include using deviations from sinusoidal variations to constrain
convective depths \citep{mikemon05} and using deviations from equally spaced
overtones to discern interior composition gradients \citep[e.g.][]
{groot10,kaw94}. In
many cases, the best results are achieved for stars with highly constrained
observations. Observational constraints can include the usual spectroscopic
measurements ($\log g$, $T_{\rm eff}$ and some compositional constraints),
number and characterization of periodicities (frequency or period, amplitude,
phase, and pulse shape) as well as frequency multiplets and equal period spacings
which associate specific periodicities with pulsation modes.

%

In brief, nonradial pulsations (periodicities) are characterized by 
three quantized numbers (modes)
$n$, $\ell$, and $m$. These represent the number of radial nodes ($n$),
surface nodes ($\ell$) and azimuthal surface nodes ($m$).
In the asymptotic limit for $n\gg\ell$, gravity ($g-$)modes should be 
equally spaced
in period for consecutive values of $n$ according to the expression:
\begin{equation}
\Pi_{\ell ,n}\,=\,\frac{\Pi_o}{\sqrt{\ell\left(\ell +1\right)}}n+\epsilon
\end{equation}
where $\Pi_o$ and $\epsilon$ are constants, in seconds 
 \citep[see][among others]{aerts10,tass,smey,unno}. 
The period spacings between two consecutive overtones are:
\begin{equation}
\Delta \Pi_{\ell}\,=\,\frac{\Pi_o}{\sqrt{\ell\left(\ell +1\right)}}
\end{equation}
where $\Delta \Pi_{\ell}\,=\,\Pi_{\ell ,n+1}-\Pi_{\ell ,n}$. Because of
geometric cancellation  \citep{me1,dzie}, 
$\ell =1$ and $ 2$ modes are the most likely
nonradial modes to be observed and the specific
relations between them are:

\begin{equation}
\Pi_{n,\ell =2}\,=\,\frac{\Pi_{n,\ell =1}}{\sqrt{3}}+C
\end{equation}
where $C$ is a constant that is expected to be small and 
is zero if $\epsilon_2=\epsilon_1$, and
\begin{equation}
\Delta\Pi_{\ell =2}\,=\,\frac{\Delta\Pi_{\ell =1}}{\sqrt{3}}.
\end{equation}
The asymptotic approximation applies to periods within completely
homogeneous stars. However real stars, particularly compact stars
for which gravitational settling is important and hot stars in which
radiative levitation is important, develop compositional discontinuities
where the mean molecular weight changes. The \emph{transition zones}
of compositional changes can work as a reflective wall which confines 
pulsations to specific stellar regions. This ``trapping'' of pulsation
modes changes the spacing between consecutive overtones compared to the
average spacing $\Delta\Pi$. Trapped modes can be used to deduce structural
changes associated with chemical transitions \citep[e.g.][]{kaw94,cost08}.

The period spacing relations are independent of $m$ and are applied under
the assumption of $m\,=\,0$ periodicities. In the case of extremely 
slowly rotating stars, this may be a valid assumption. However for stars 
which complete several revolutions within a set of observations, pulsations
will create frequency multiplets.  To first order, 
these multiplets will have $2\ell +1$ components spaced at 
$$\nu_{n,\ell ,m}\,=\,\nu_{n,\ell,0}\,+\, m\Omega\left( 1-C_{n,\ell}\right)$$
where $\Omega$ is the rotation frequency and $C_{n,\ell}$ is the Ledoux
constant \citep{led51}.

In this paper, we apply Eqns.~1 to 4 to $g-$mode pulsations observed
in hot subdwarf (sdB) variables.
Subdwarf B variables were first discovered
in 1996 and now consist of two well-established classes. These are the 
short-period pressure ($p-$)mode pulsators which are designated 
V361~Hya stars \citep{kill97} and longer-period gravity ($g-$)mode pulsators
designated V1093~Her stars \citep{grn03}. There are also hybrid pulsators,
sometimes called DW~Lyn stars after that prototype \citep{schuh06}, which
show both types of variations. About 50 V361~Hya pulsators have been 
detected \citep{roysurv} with a couple dozen receiving various amounts
of follow-up data \citep[see for example ][]{reed07b}. However, observational
constraints on pulsation modes are extremely rare for the V361~Hya class,
occurring only twice using multiplets \cite{me2,andy09}. Time-resolved
spectroscopy, sometimes coupled with multicolor photometry,
has had some limited success \citep{to04,to06,reed09,andy10}.
See \S 2.2 through 2.4 of \citet{royreview} for a recent review of
these methods. The lack of observational constraints has led
model-matching efforts to proceed by using the forward
method, which consists of matching observed
periods to those of models, with the closest fit, within spectroscopic
constraints, being deemed the correct one \citep[For a review see ][]
{char09}. Progress on $g-$mode pulsators  has been slow because of the 
difficulties in observing many pulsation cycles for 
periodicites of one to three hours in extremely blue stars from the ground.
With the acquisition of long time-series photometric data from satellites,
such as \emph{Kepler} and CoRoT, detailed asteroseismology of 
the V1093~Her pulsators is now possible. 

The \emph{Kepler} spacecraft has a primary mission to find Earth-sized
planets within the habitability zone around Sun-like stars \citep{bor10}.
To do this, the spacecraft continuously examines roughly 150,000 stars
in search of transits. As a byproduct of that search, high quality 
photometric observations are obtained which have proven extremely useful
for the study of variable stars \citep{koch10,prsa10}. 
The \emph{Kepler} spacecraft has two effective integration times: A short
cadence (SC) integration near 1 minute and a long cadence (LC) integration
near 30 minutes.
The first year of the \emph{Kepler} mission was dedicated to a survey phase
where many target buffers were assigned to SC observations, which switched
targets on a monthly basis \citep{jenk10}. Papers~I through VII 
\citep{royP1,royP6,sdBVP1,lpsdBVbinP1,groot,reed10,andyP7} of this series along with
\citet{roy2M} describe the
search and resulting detections of periodicities in compact stars from
\emph{Kepler} survey phase observations. Papers~I and VI provide a
spectroscopic analysis and pulsation
overview of all compact stars observed with \emph{Kepler};
Papers~II, III, V, VII, and \citet{roy2M} describe the 
specific pulsation periods of the pulsators and Paper~IV associates a 
model with one pulsator using the Forward Method. Those papers serve as a 
complete introduction to pulsating sdB stars using \emph{Kepler}
SC observations.  Additionally,
CoRoT (COnvection, ROtation, and planetary Transits
satellite)\footnote{http://smsc.cnes.fr/COROT/} has observed one V1093~Her
pulsator; KPD~0629-0016 
\citep[hereafter KPD~0629;][]{charp10}.

\section{Detection and significance of regular period spacings}
In Paper~III \citep{reed10} we identified 26 of 27 periodicities for
KIC10670103 as $\ell =1$ or 2 using the relations of Eqns.~2 through 4.
In this section we search all V1093~Her stars with space-based
(13 \emph{Kepler} and 1 CoRoT) observations and apply significance
tests.
Basic information for the 14 stars of this study are provided in
Table~\ref{tab01}. This includes
 Kepler Input
Catalog (KIC) numbers, stellar designations from other sources, 
and spectroscopic properties from Papers~I and VI 
\citep[except for KPD~0629, which are from][]{charp10}. 

\begin{table}
\caption{Properties of the stars in this paper. Columns 1 and 2 supply the 
Kepler Input Catalog number and a more common name, column 3 lists
the observing period in quarter and month, columns 4 and 5 
provide the \emph{Kepler} magnitudes ($K_p$) and estimated contamination 
factors ($F_{cont}$),
columns 6 and 7 supply spectroscopic parameters, column 8 lists if the star is
in a known binary (RE = reflection effect binary; EB = eclipsing binary; EV
= ellipsoidal variable binary; and
the inferred companion is in parentheses), and column 9 lists references from this series of
papers. RH\O ~ indicates \citet{roy2M} and CH10 indicates \citet{charp10}.\label{tab01}}
\begin{tabular}{rcccccccc}
\hline
KIC & Name & Q & $K_p$ & $F_{cont}$ & $T_{\rm eff}$ & $\log g$ &  Binary &
Ref \\ \hline
2697388 & J19091+3756 & 2.3 & 15.39 & 0.149 & 23.9(3) & 5.32(3) &  - & I,III\\
2991403 & J19272+3808 & 1   & 17.14 & 0.601 & 27.3(2) & 5.43(3) & RE(dM)& I,V\\
3527751 & J19036+3836 & 2.3 & 14.86 & 0.081 & 27.9(2) & 5.37(9) &  - & I,III\\
5807616 & KPD~1943+4058&2.3 & 15.02 & 0.332 & 27.1(2) & 5.51(2) &  - & I,III,IV\\
7664467 & J18561+4319 & 2.3 & 16.45 & 0.879 & 26.8(5) & 5.17(8) &  - & I,III\\
7668647 & FBS1903+432 & 3.1 & 15.40 & 0.226 & 27.7(3) & 5.45(4) &  - & VI,VII \\
8302197 & J19310+4413 & 3.3 & 16.43 & 0.256 & 26.4(3) & 5.32(4) &  - & VI,VII \\
9472174  & 2M1938+4603 & 0  & 12.26 & 0.022 & 29.6(1) & 5.42(1) & EB(dM) &I,RH\O\\
10001893 & J19095+4659 & 3.2 & 15.85 & 0.710 & 26.7(3) & 5.30(4) &  - & VI,VII \\
10553698 & J19531+4743 & 4.1 & 15.13 & 0.385 & 27.6(4) & 5.33(5) &  - & VI,VII \\
10670103 & J19346+4758 & 2.3 & 16.53 & 0.450 & 20.9(3) & 5.11(4) & EV(WD)& I,III\\
11179657 & J19023+4850 & 2.3 & 17.06 & 0.129 & 26.0(8) & 5.14(13) &  RE(dM)&I,V\\
11558725 & J19265+4930 & 3.3 & 14.95 & 0.028 & 27.4(2) & 5.37(3) &  - & VI,VII\\
- & KPD0629-0016 & - & 14.91$^{\dagger}$ & - & 27.8(3) & 5.53(4) &  - & CH10  \\ \hline
\end{tabular}
\end{table}

\begin{figure}
 \centerline{\psfig{figure=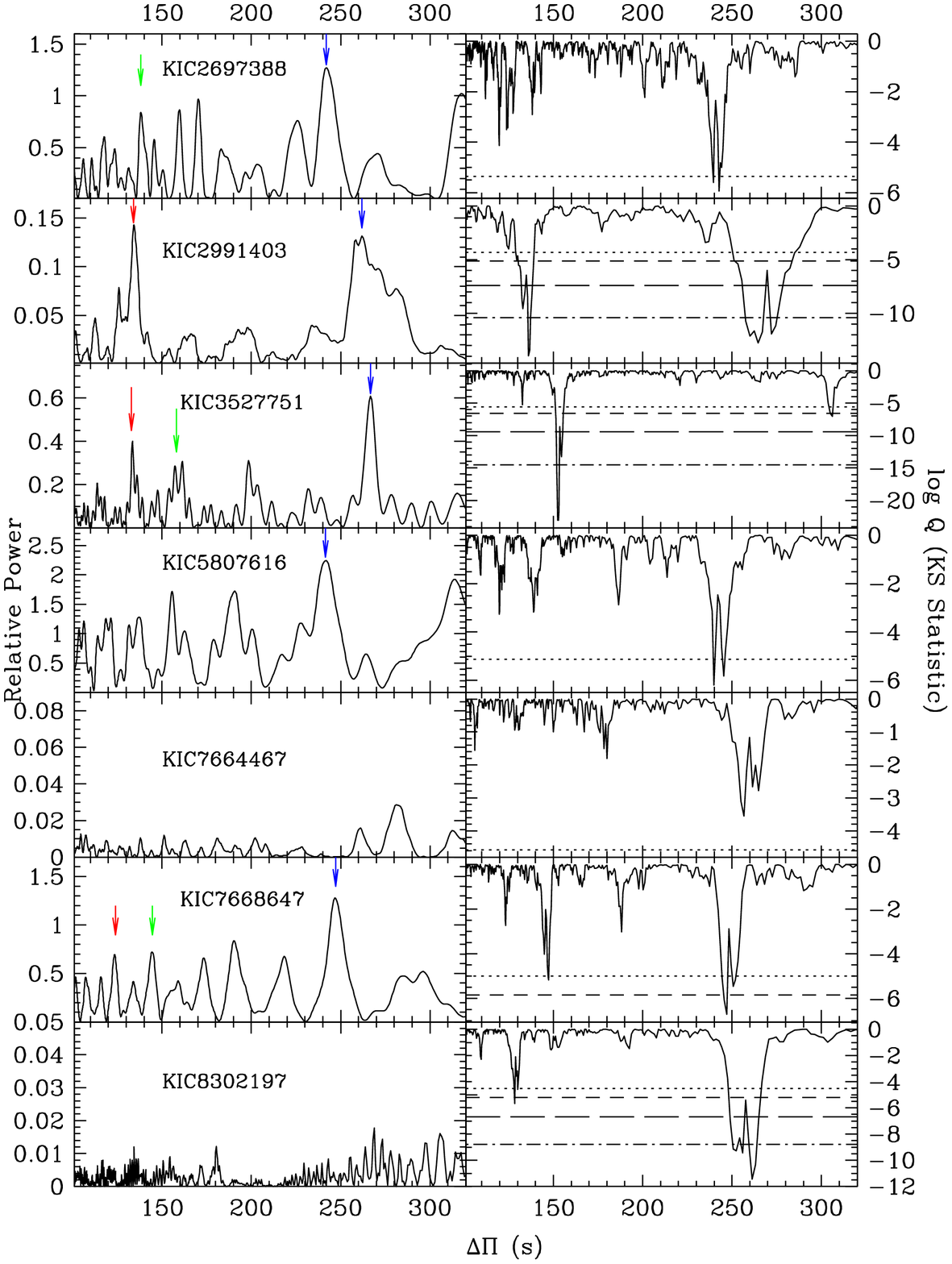,width=3.7in}\psfig{figure=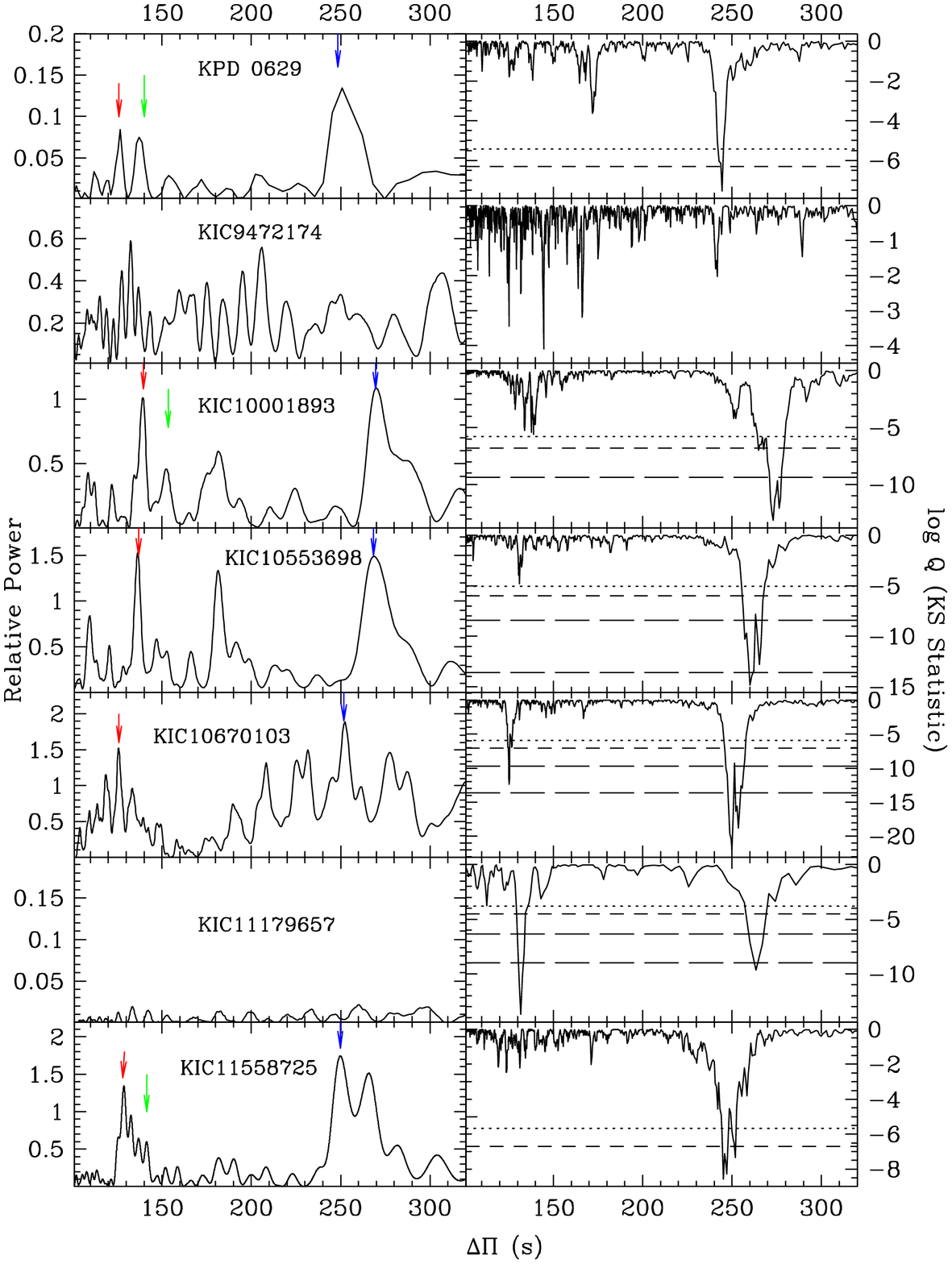,width=3.7in} }
\caption{Left panels: Indications of regular period spacings using Period transforms. Blue arrows indicate the 
$\ell =1$ spacings, green arrows the $\ell =2$ spacings, and red arrows aliases. Right panels:
The Kolmogorov-Smirnov (KS) test applied to the detected periodicities. Confidence levels
of 90\% (dotted line), 95\% (short dashed line), 99\% (long dashed line), and 99.9\% (dot-dashed
line) are shown. These confidence levels are calculated for the range
of period spacings between 100 and 400 seconds, assuming uniformly
distributed random periods.} \label{fig01}
\end{figure}

From our work with KIC10670103, we were expecting $\ell =1$ period
spacings near to 250~s and there are several other stars (particularly KIC8302197
and KIC10001893) which trivially show equal period spacings (or multiples thereof) very near
to this value. Since V1093~Her stars have small ranges for $T_{\rm eff}$
and $\log g$, we anticipated that all our targets should have $\ell =1$ period
spacings near to 250~s.
We took the dual approach of \citet{wing91}, to search for regular period spacings using
period transforms (PT) and Kolmogorov-Smirnov (KS) tests.
The period transform is an unbiased test where power spectra are converted to period
spectra and then a Fourier transform is taken of that. Peaks in the PT indicate
common period spacings. We used
the $g-$mode region from $0-1000\,\mu$Hz from our power spectra.
The PT method is 
sensitive to the number of periods from which to find correlations. \citet{wing91} did this
for the pulsating DOV star PG~1159-035 (also known as GW~Vir), 
for which 125 periods were detected.
Conversely, our richest $g-$mode pulsator only has 46 periodicities and our poorest a
meagre seven. As such, our expectations were low and we were happily surprised by the 
success of this method. For 10 stars (shown in the
left panels of Fig.~\ref{fig01}) the $\ell =1$ 
peak corresponding to a regular period spacing near 250~s is readily picked out 
and in five of these, we can deduce the 
$\ell =2$ peak as well using Eqn.~4. We then fitted the PT with 
a nonlinear least-squares technique to determine
the period spacing values and errors for each one. 
For KIC3527751, an alias occurs for $\Delta\Pi_1+\Delta\Pi_2$
and KIC11558725's peaks are split because of small period spacings (possibly related
to rotational multiplets). KIC8302197 and KIC9472174 do not have any peaks that stand
out. For KIC8302197, this most likely occurs because of the few periods (9) and
for KIC9472174 this is likely related to the short data series (9.7~d) and the 
complexity within the FT caused by binarity.

The Kolmogorov-Smirnov (KS) test is a nonparametric test that compares a sample 
distribution [$F_n(x)$] with a reference distribution \citep[Eqn.~1 in our 
case;][]{chak}. The KS test has proven useful with white dwarf 
pulsators \citep{wing91,kaw88}.
The KS test uses previously detected pulsation periods as input and
so has a selection effect caused by our detections. Any such effect should be small
as the data
are nearly gap-free, and so period detections should be accurate. However,
some stars show small, marginally-unresolved periodicities and these could
skew the results as they are sometimes included and other times excluded in
the period lists.
We applied the KS test for equal period spacings
between 50 and 800 seconds. The results for the range of 100 -- 300~s 
are shown in the right panels of 
Fig.~\ref{fig01}. Unlike the PT test, the KS test has a local minimum for
all stars for period spacings near 250~s, except for KIC3527751, where it detects
spacings of 152~s. The PT test for KIC3527751 
shows both the $\ell =1$ and 2 peaks, with the $\ell =1$ having a
higher amplitude, whereas the KS test only detects an extremely strong $\ell =2$ spacing. 
The $\ell =2$ spacing is within the errors of the both the PT test and a linear
least-squares fit
and match the value determined using Eqn~4 based on our $\ell =1$ determinations.
However, only for the KS test does it dominate.
The $\Delta\Pi =241$~s detection for KIC9472174 is not significant.
KIC9472174 is a fairly rich pulsator with a large range of periods and in
a known short period binary. There are many small spacings between 100 and 
150~s and it could be that many of these are parts of rotationally split
multiplets. With a binary period of 3.0~h, any rotational multiplets would
have splittings of $\approx 92\,\mu$Hz, which means they would overlap
asymptotic spacings and disrupt our period spacing detection techniques.
While the PT test was not useful for KIC8302197 as it has too few periods, 
the KS test readily found a regular period spacing. All of the 
KS results show some quantity of multiple peaks caused by deviations in the period
spacings. In PG~1159, these are attributed to mode trapping. Stars which show weak
secondary peaks indicative of $\ell =2$ spacings 
include KICs 2991403, 5807616, 7664467,
7668647, 10553698, 11558725, and KPD~0629. Surprisingly, the $\ell =2$ spacings
for KIC10670103 produce an insignificant peak, 
even though we previously detected eight $\ell =2$ modes including five
consecutive overtones.

Using the period spacings found in the PT and KS tests (or integer multiples thereof), 
we identified periods as
$\ell =1$ or 2, or unknown\footnote{Periods and mode identifications appear in
Tables~4 through 17 of the on-line appendix.}. 
We then did a least-squares straight line fit to each
$\ell =1$ or 2 series, arbitrarily assigning $n$ values such that $n$ was not
negative\footnote{Except for KIC5807616, where $n$ is chosen to match Paper~IV.}
and satisfying Eqn.~3 between the $\ell =1$ and 2 series. The period spacings found
using all three methods were in agreement and in Table~2 we use those from the 
linear least-squares fits, for which the errors are the most straightforward.

\subsection{Monte Carlo tests}
From the PT and KS tests, we already have strong evidence that
nearly all of these stars have regular period spacings. However,
according to stellar models \citep[][hereafter CH02]{char02}, 
even period spacings are not anticipated.
Since most of the pulsators
have rich pulsation spectra, it is reasonable to question if the detections
are just chance alignments. As a third test,
we produced Monte Carlo simulations that randomly select periods
to match with asymptotic sequences to within the errors. 
A number of observed periodicities, $N$, set to match what is observed,
were randomly selected
to fit within an observed range $P_{\rm min}\leq P \leq P_{\rm max}$.
For KIC8302197 and KIC7664467, we compared the $N$ randomly selected
periods with a single sequence of the form 
$P_n = \left( P_{\rm min}-j\delta\right) +n\times \Delta\Pi \pm \sigma$.
The quantity $\delta$ represents a small shift of the zero point which was
repeated $j$ times until $P_0\leq P_{\rm min}-\Delta\Pi$. The $j$ value 
that produced the greatest number of 
matches was used and the $n$ values were tested to see how many consecutive
overtones were detected. $\sigma$ is the difference allowed between
the random and sequence periods and is chosen to be 
slightly bigger than what is observed.
Choosing this value for $\sigma$ makes our Monte Carlo simulations extremely
conservative and the real probability of matching is much smaller since
this $\sigma$ is typically valid for only one or two periods, with most
errors being much smaller. $n$ is then stepped until $P_n >P_{\rm max}$. 
To generate an $\ell =2$ appropriate sequence, we simply divided the
$\ell =1$ $P_n$ by $\sqrt{3}$, as required by Eqn.~3 
and then extended $P_{\rm max}$
so even the longest random periods could have $\ell =2$ matches.
Our code also insured that only one randomly selected
period matched each possible sequence period (eliminating double counting).
A million sets of random periods were generated in each Monte Carlo
simulation and the resulting matches
were converted to percentages in Column~9 of Table~\ref{tab02}.

\begin{table}
\caption{Period spacings determined from linear least-squares fits.
Column 1 provides the KIC number (KPD designation for the CoRoT star),
Columns 2 and 3 are the $\ell =1$ and 2 period spacings (errors
in parentheses), columns 4, 5, 6, 7, and 8 provide the total number of periods,
the number assigned as $\ell =1$ and 2, and  the number of consecutive $\ell =1$
and 2 overtones. Parenthetic numbers in Column 6 indicate the number of modes which
are ambiguous between $\ell =1$ and 2 identifications. They are not counted as
$\ell =2$ in Column 6. The last column provides the percentage of Monte Carlo 
simulations that produced a match to the observations. Notes:
a) Using just the $\ell =1$ sequence with five consecutive overtones.
b) Leaving three deviant $\ell =2$ matches as unassigned.
c) 10 million simulations produced no results which included 13 consecutive overtones.
d) Assuming f1 is $\ell =1$. 
e) Periods f1 and f2 are counted as $\ell =2$.
f) Leaving the deviant periods f17 and f44
unassigned and assuming f2 is $\ell =2$.
\label{tab02}}
\begin{tabular}{rcccccccc}
\hline
Star & $\Delta \Pi_1$ & $\Delta \Pi_2$ & N & $N_1$ & $N_2$ & $NC_1$ & $NC_2$ & MC\% \\ \hline
2697388 &  240.07 (0.27) & 138.54 (0.16)  & 36 & 16 & 13 (2) & 4  & 3 & 0.04 \\
2991403 &  268.52 (0.74) & 153.84 (1.19)  & 16 & 7 & 4       &  5 & 0 & 0.009$^a$ \\
3527751 &  266.10 (0.38) & 153.57 (0.12)  &38 & 15 & 14 (5) & 2  & 2 & 0.018 \\
5807616 &  242.12 (0.62) & 139.13 (0.38) & 22 & 11 & 6 (3)  & 3  & 3 & 23.0 \\
7664467 &  260.02 (0.77) & -             & 7 & 6  & 1      & 0  & -  & 0.16\\
7668647 &  248.15 (0.44) & 144.71 (0.57) & 18 & 12 & 5 (2)  & 2  & 0 & 0.0014 \\
8302197 &  257.70 (0.56) & -             & 9& 9  & -      & 2  & -  & 0.0007 \\
9472174  & 255.63 (0.30) & 147.70 (0.69) & 20 & 8 & 8 (1)   & 2  & 2  & 5.4$^b$ \\
10001893 & 268.53 (0.61) & 154.74 (0.34) & 26 & 18 & 9 & 12 & 3 & 0.0$^c$ \\
10553698 & 271.15 (0.54) & 156.68 (0.31) & 30 & 12 & 9 (6)  & 6 & 3  & 0.22 \\
10670103 & 251.13 (0.31) & 145.59 (0.26) & 27 & 19 & 8  & 5  & 5  & 0.04 \\
11179657 & 231.02 (0.02) & 133.64 (0.40) & 12 & 3  & 7 (1)  & 0 & 3 & 0.0002$^d$ \\
11558725 & 246.77 (0.58) & 142.57 (0.14) & 46 & 18$^e$ & 13 (1) & 8 & 2  & 2.2$^f$ \\
KPD0629 &  247.17 (0.48) & 142.74 (0.30) & 17 & 12 & 3 (2) & 3  & 2  & 1.1 \\ \hline
\end{tabular}
\end{table}

The Monte Carlo simulations indicate that 10 stars have less than a 1\% chance that
their regular period spacings are the product of random chance. KPD~0629 has a 1.1\% 
chance and KIC11558725 has a 2.2\% chance of occurring randomly from our Monte
Carlo simulations. KIC5807616 and KIC9472174 are unconstrained from this
test.

\begin{table}
\caption{Sample table of supplemental material. Tables for all 14 stars appear
on-line.
Periods and period spacings for KIC2697388.
Identifications (column 1) and periods are those from Paper~III \citep{reed10}.
Columns 3, 4, and 5 provide the mode degree $\ell$ and the overtone fit
to $P_{\ell}=P_{\ell o}+n\cdot\Delta 
P_{\ell}$ where n is arbitrarily chosen such that there are no negative values,
except for KIC5807616, where it is chosen to match Paper~IV.
Column 6 provides the difference between the observed and asymptotic
relation period, column 7 lists the fractional period differences and
column 8 is the observed spacing $P(n_{\ell ,i})-P(n_{\ell,j})/(i-j)$. 
It is ambiguous whether $\ell =1$ or 2 modes should be associated with f25,
f23, and f11. f25 was not used for the $\ell =1$ fit as it is most likely
$\ell =2$.
\label{tab03}}
\begin{tabular}{lccccccc}
\hline
ID & Period  & $\ell$ & n$_1$ & n$_2$ & $\delta P$ & $\delta P/\Delta P$ & Spacing  \\
 & (sec) &  & & & (sec) &  &  (sec)\\ \hline
f30& 2757.118 & 2 & -  & 15& 19.393 &  0.140 & 140.7 \\
f29& 3008.732 & 2 & -  & 17&-6.072  & 0.044 & 125.8 \\
f28& 3517.219 & 1 & 10 & - & -8.790 & -0.037 & 254.2 \\
f27& 3700.224 & 2 & -  & 22& -7.277 & -0.053 & 138.3 \\
f26& 3757.261 & 1 & 11 & - & -8.815 & -0.037 & 240.0 \\ 
 . &     .    & . & .  & . &    .   &   .    &    .  \\ 
 . &     .    & . & .  & . &    .   &   .    &    .  \\ 
 . &     .    & . & .  & . &    .   &   .    &    .  \\ \hline
\end{tabular}
\end{table}

\section{Ensemble and Model Comparison}
\begin{figure}
 \centerline{\psfig{figure=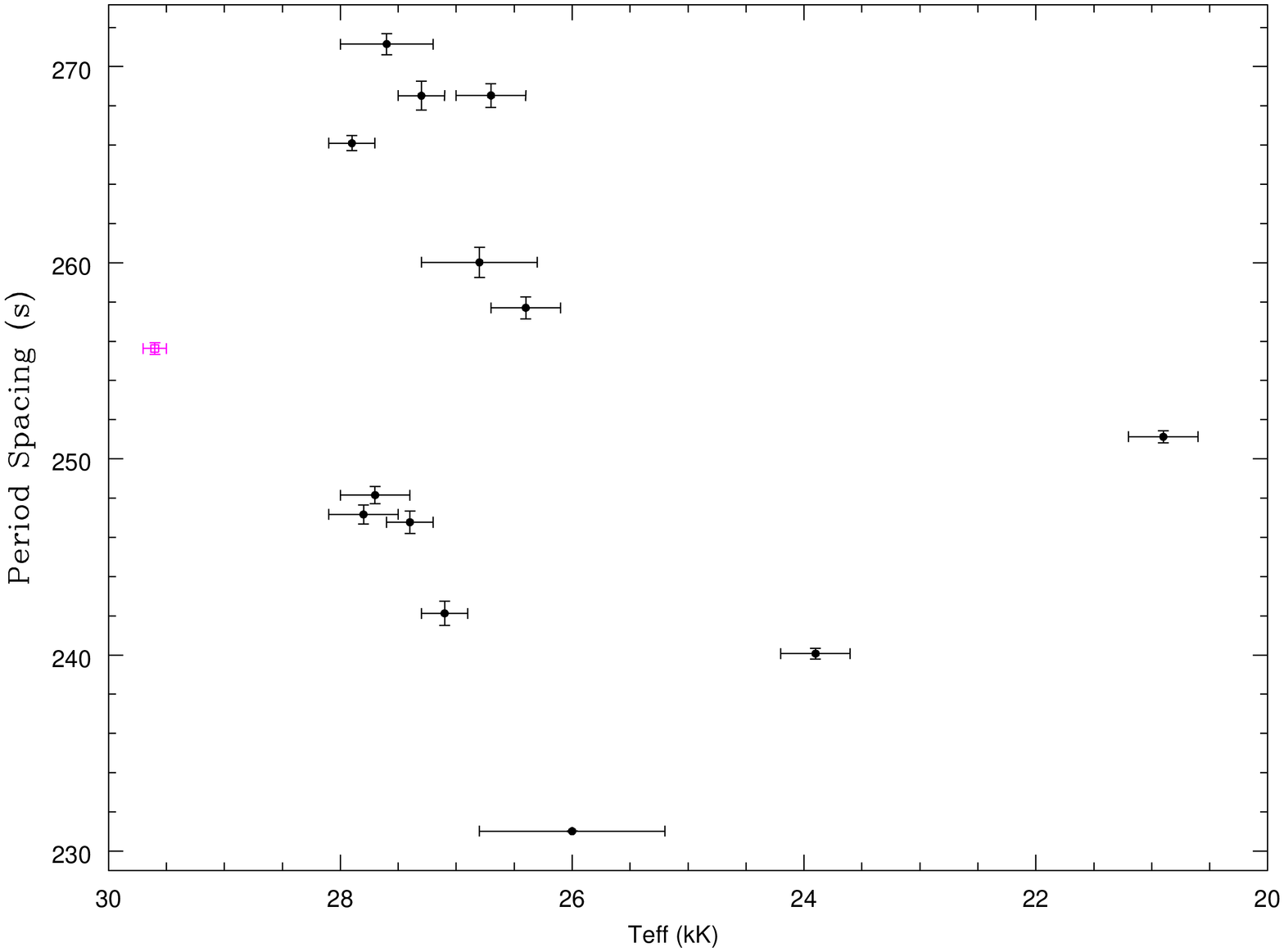,angle=-90,width=3.7in}\psfig{figure=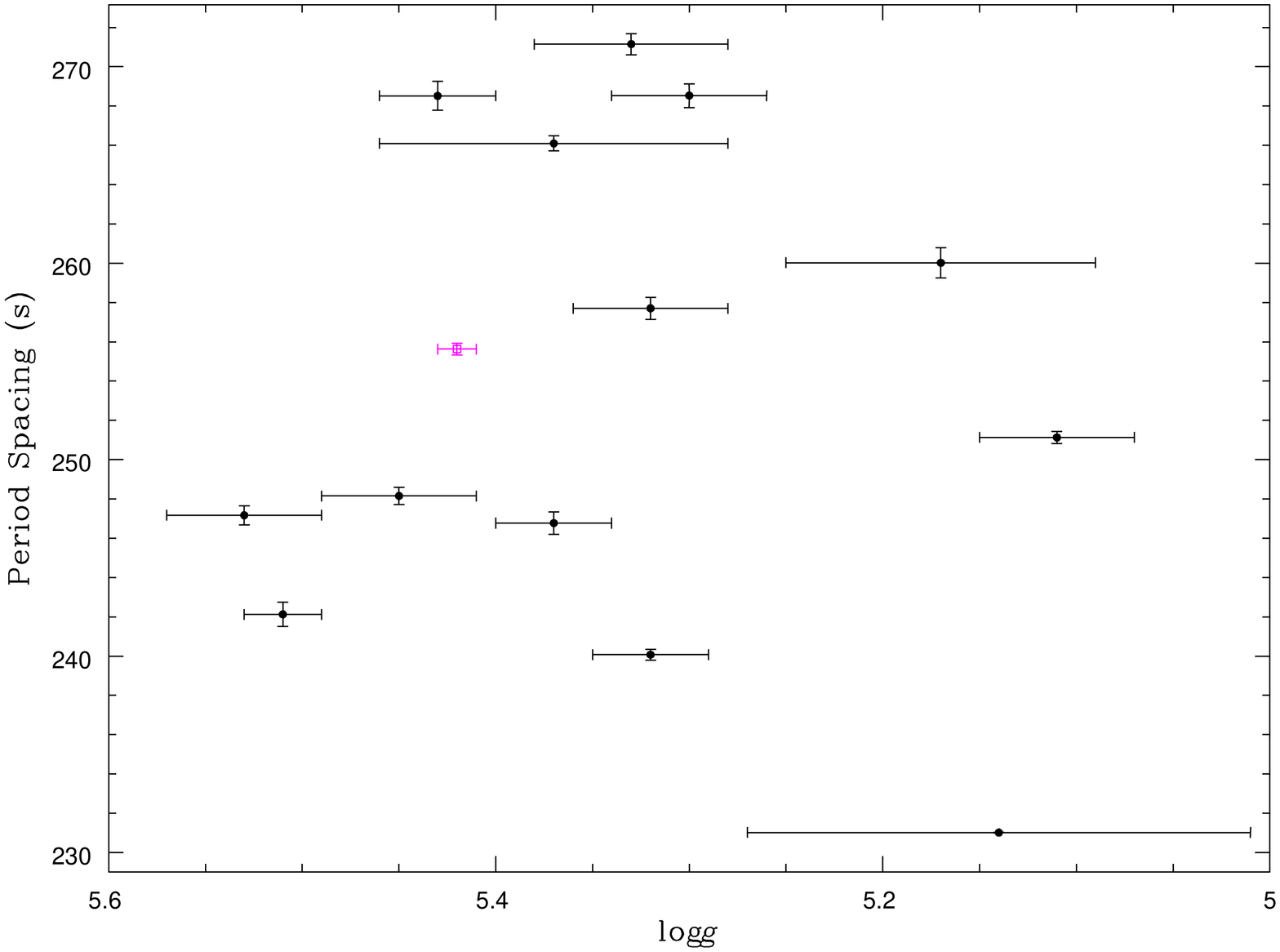,angle=-90,width=3.7in}}
\caption{Period spacings compared with $T_{\rm eff}$ and $\log g$. The 
red point indicates KIC9472174, which is the only star for which PT, KS, and
MC tests were all inconclusive.
} \label{fig02}
\end{figure}

Figure~\ref{fig02} shows our detected $\ell =1$ period spacings with gravity and
effective temperature. The temperatures span nearly 10~000~K while $\log g$
only covers 0.6~dex. Naturally, what is sought is
a relationship between period spacings and physical properties, such as
Eqn.~5 of \citet{kaw94} for white dwarfs. Such a relationship would allow
the determination of properties based on period spacings alone. While
white dwarfs and sdB stars are both compact stars, there is no \emph{a priori}
reason to expect that any correlations should exist for sdB stars.
Figure~31 of CH02 indicates that as
envelope thickness decreases, the distance between trapped modes increases as
do period spacings. However, the effect of trapped modes is increased with
decreasing envelope thickness, and so while there are more overtones between
trapped modes, the impact of a trapped mode would be to eliminate any
sequence of the form of Eqn.~1 longer than three or four consecutive periods.
Figure~16 of CH02 indicates that the longest period spacings should
occur where $T_{\rm eff}$ and $\log g$ are both small or both large, though
they only test for $g-$modes with $n\leq 9$, which may be too small for asymptotic
relations. 
However, in Fig~\ref{fig02} there do not appear to be any
trends, either with gravity or temperature. Since 10 stars have temperatures
near 27~500~K yet period spacings that range from 242 to 271~s while the
extremophiles of the group have period spacings near the middle of this
range, it would have
to be deduced that temperature does not impact period spacings in sdB stars.
No trends are obvious with $\log g$ either, though in
this case the span is much smaller compared with the associated errors.
Table~3 of CH02 indicates that period spacing should increase
with decreasing envelope mass. It would useful to compare the CH02
models with Paper~IV, but unfortunately the CH02 paper calculates
for $\ell =3$ modes and Paper~IV does not, making a direct comparison difficult.
Appropriate stellar models will have to be produced to determine what the 
parameter(s) is (are) that affects the period spacings, but this paper is concerned
with interpreting observations and so we will not address modeling issues.

\begin{figure}
 \centerline{\psfig{figure=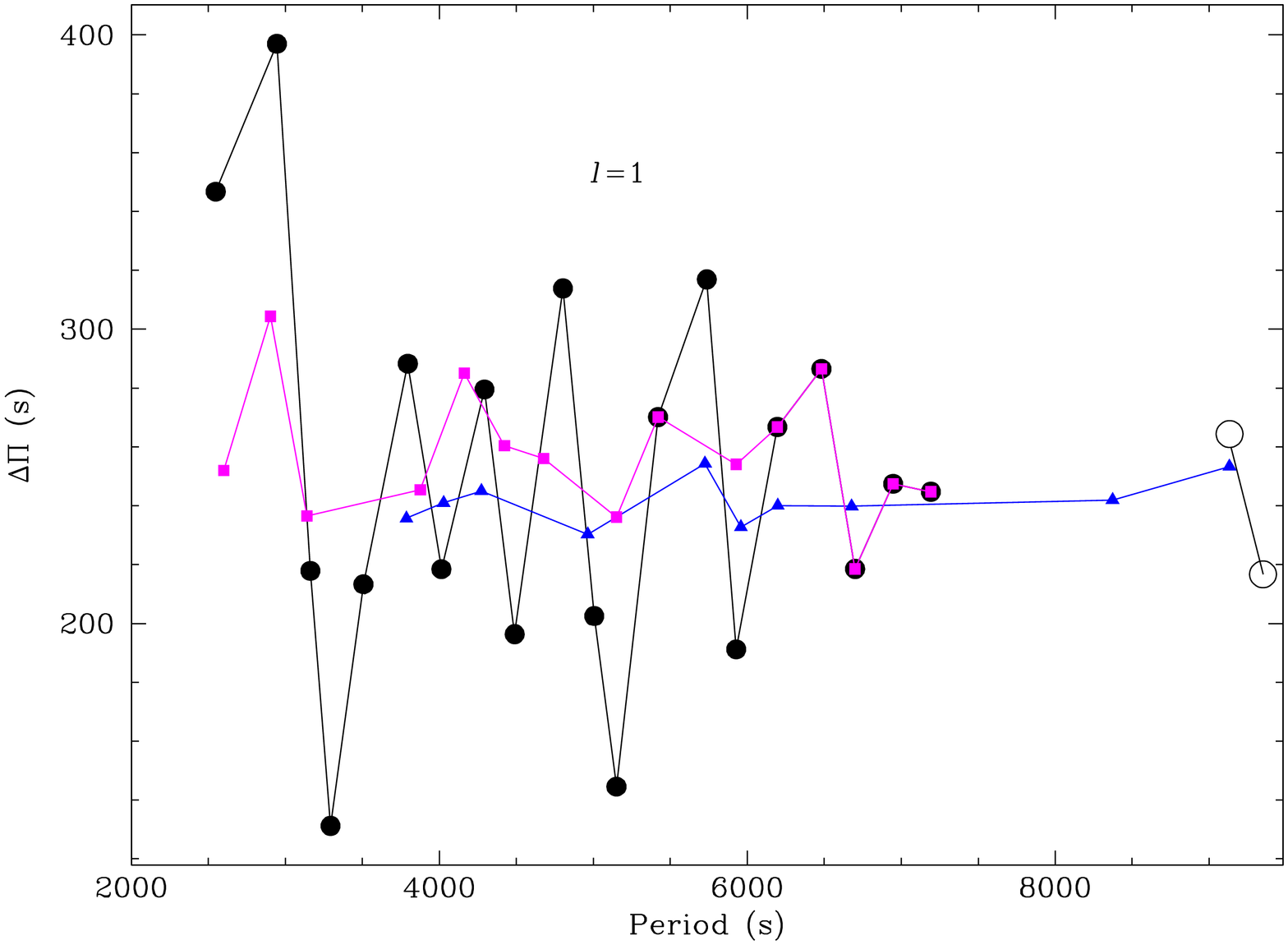,angle=-90,width=3.7in}\psfig{figure=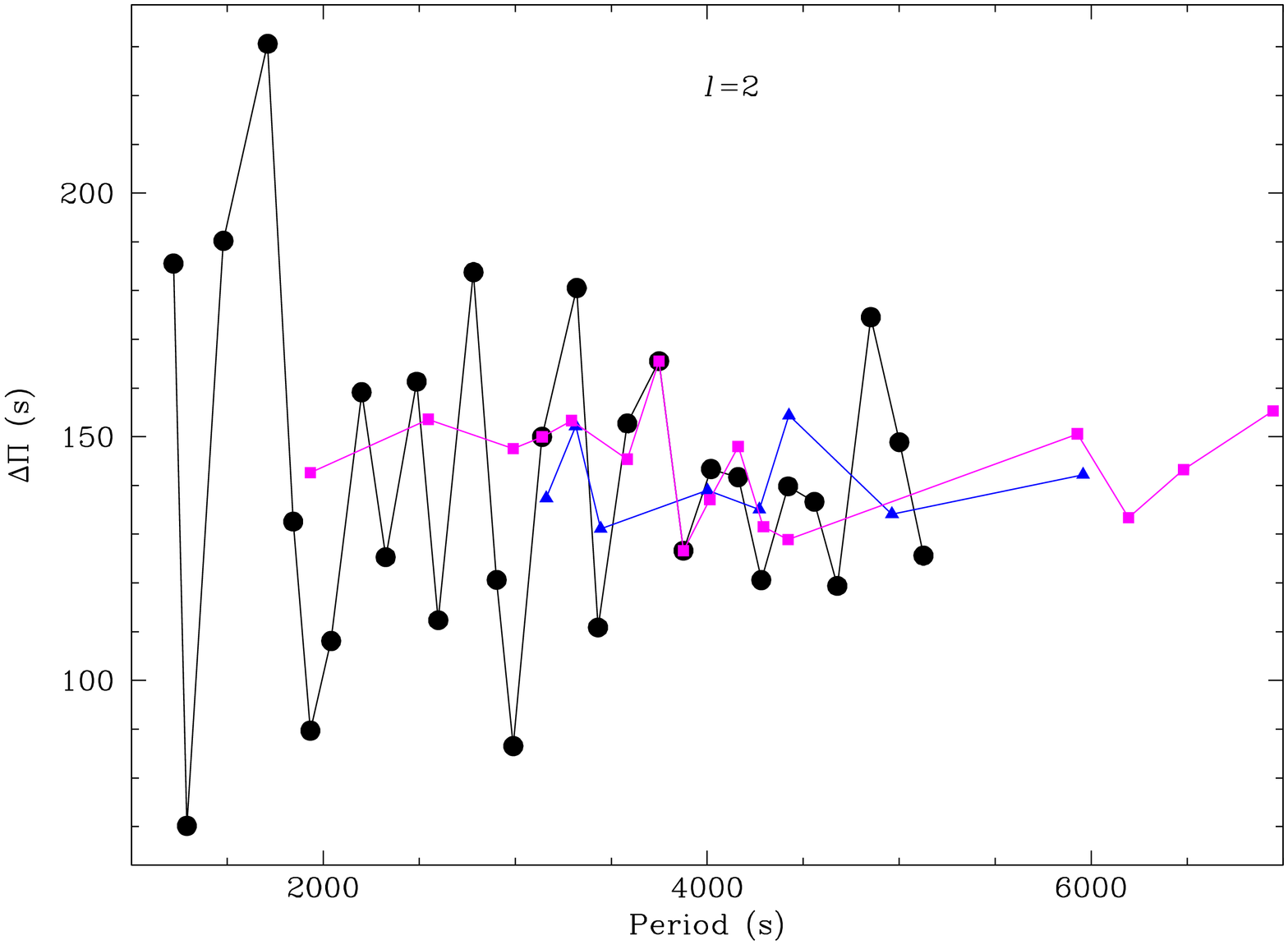,angle=-90,width=3.7in}}
\caption{A comparison between model, observed, and linearly fit model data for
KIC5807616. Black circles are model periods from Paper~IV
(open circles were not used in the fit), blue triangles are those
found from \emph{Kepler} data and magenta squares are those found from
a blind fit using the observed period spacing with model periods. If a 
magenta square is plotted over a black circle, then our blind fit matched
the model's mode assignments.
} \label{fig03}
\end{figure}

\begin{figure}
 \centerline{\psfig{figure=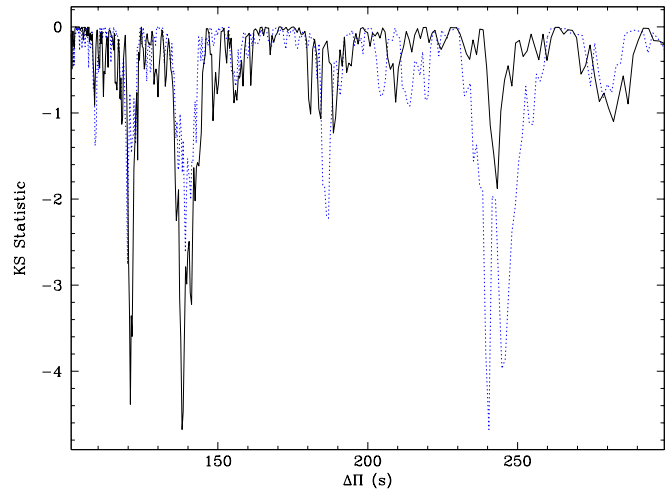,angle=-90,width=3.7in}}
\caption{The KS test applied to 51 model periods from Paper~IV.
The dotted (blue) line is the KS test for KIC5807616 from
Fig.~\ref{fig01}.
} \label{fig04}
\end{figure}

While CH02 examined period spacings for gravity (and pressure) modes,
the model they used was significantly hotter than these stars. The model of
Paper~IV is obviously appropriate as it was made
to match KIC5807616 and so we compared it with our findings. Figure~\ref{fig03}
shows the model spacings for many of the $\ell =1$ and 2 modes (black
circles). The
$\ell =1$ period spacings range from $\approx 50$ to 400~s with mode
trapping dominating the spacings. In the sequence of 21 period spacings, 
only twice is the change between consecutive spacings smaller than 20~s while
the rest are greater than 50~s. For comparison, the period spacings we
selected for KIC5807616 (which changed by less than 25~s for all $\ell =1$
modes) are shown as blue triangles. Naturally, one could
pick out just the peaks or troughs of the model and get more consistent period
spacings that way, but you would only rarely get a sequence of three
consecutive overtones. To test this assumption we performed a blind test on 51
model $\ell =1$ and 2 periods from Paper~IV; including model
sequences of 21 consecutive $\ell =1$ and 30
$\ell =2$ modes. Putting them in period order only (removing the model mode
assignments) and using the observed period spacings as a guide, we assigned
periods as $\ell =1$ or 2, or left them unassigned.
Allowing periods to deviate by up to 32~s from equal spacings (28\% more than
the observed deviations), we assigned 15 $\ell =1$
and 16 $\ell =2$ modes (double counting eight periods, which were ambiguous
between the modes). Of the 15 $\ell =1$ assignments, eight were model 
$\ell =1$ modes and of the 16 possible $\ell =2$ mode assignments, eight were
model $\ell =2$ modes (two others were close). Our mode assignments from the model
periods are shown as (magenta) squares in 
Fig.~\ref{fig03}. When squares are plotted over circles, our blind test mode
assignments match those of the model. 
As expected, this test indicates that mode identifications
using equal period spacings does not work well if there
is any significant mode trapping since Eqn.~1 biases us to selecting periods with
small (or no) mode trapping. We also applied the KS test to the 
51 model periods and the
results are shown in Fig.~\ref{fig04}. The KS test preferentially 
detects $\ell =2$ period
spacings with a mild $\ell =1$ period spacing. For comparison, the KS test for 
KIC5807616's observed periods is shown as a 
dotted (blue) line and shows that the actual data has
much stronger $\ell =1$ period spacings. However, the period spacings detected
in the models are about correct, indicating that perhaps with more subtle
mode trapping, the model would better approximate the observations. In Table~6 of
the supplemental material, we show our mode assignments as well as those of
Paper~IV. We chose the radial order $n$ to match the model at f11=4027~s. When mode
assignments via regular period spacings and those from the model agreed, so did the
radial order. Four of our 11 $\ell =1$ mode assignments matched those of the
model and seven of our nine $\ell =2$ mode assignments matched. Again, this
likely indicates that the star does not trap modes as significantly as the
model predicts. Since this paper is concerned with observed mode identifications
and period spacings, we leave a detailed model analysis to those best suited
to do them.

\section{Summary}
We tested 13 \emph{Kepler}-observed and one CoRoT-observed 
$g-$mode pulsating subdwarf B stars for consistent period spacings
which can be used to observationally identify pulsation modes. 
We used two different spacings detection tests, a period transform (PT)
and a Kolmogorov-Smirnov (KS) test and a Monte Carlo (MC) significance test.
The PT test identified 10 stars
as having consistent $\ell =1$ period spacings and five
of these also showed indications of $\ell =2$ period spacings. Our KS test
results clearly detected $\ell =1$ constant
 period spacings in \emph{all} our program
stars, except KIC9472174, which has a period spectrum complicated by 
binarity (though the KS result does have an appropriate local minimum) and
KIC3527751, for which it finds a very strong $\ell =2$ period spacing.
A further five stars show local minima appropriate for $\ell =2$ period 
spacings from their KS tests.
Monte Carlo tests indicate that for 10 stars, our mode assignments
(provided in accompanying on-line material)
are very likely correct. For three additional stars, a random cause for the
spacings is below 6\% and KIC5807616 has a 23\% chance that the equal period
spacings are being created
randomly, as a worst-case scenario. 

For all sample stars, except KIC9472174,
three of the four methods (PT, KS, MC, and linear least-squares)
find evidence for regular period spacings. 
12 of the 14 stars have $\ell =1$ and 2 period spacings which satisfy
Eqn.~4 and 11 stars have periods (45 periods in total) that satisfy 
Eqn.~3, with $C$ equal to zero. 
Combined, these provide a strong indicator that we are correctly
identifying periodicities as $\ell =1$ and 2 modes, rather than
higher degrees which have different relations. 
For these 14 stars, we assigned a total of 222
of a possible 317 periodicities as $\ell =1$ or 2 modes. Such a 
large quantity of observationally constrained modes should prove
exceedingly useful for stellar modeling.

Our results
clearly show the value of long-duration space-based observations.
While there have been some remarkable ground-based efforts to observe 
$g-$mode sdB stars, they have not resulted in sufficient
detections to evaluate period spacings. Additionally, the \emph{Kepler}
results are solely from the survey phase of the mission. Longer duration
observations should detect more pulsation periods, including higher degree
 ($\ell\geq 3$) modes, which we have not searched for at all.

\section{conclusion}
In order for Eqn~1 to be useful, mode trapping must be small (or none).
Since Eqn~1 produced a large fraction of significant mode assignments
for nearly all of the stars we examined, mode trapping must be 
substantially reduced from what current models indicate.
Figure~16 of CH02 shows period spacings against both $T_{\rm eff}$
and $\log g$ for $g-$mode pulsations. Unfortunately, it only has $n\leq 9$,
where evenly-spaced periods are not expected. However, for higher $n$ values,
such a plot should show a flat surface. According to CH02,
$\Delta \Pi$ shows a plateau of maximum values running from the lowest
 $T_{\rm eff}$ and $\log g$ to the highest $T_{\rm eff}$ and $\log g$.
However those results are significantly affected by mode trapping, and so
may not be a clear indicator of trends in period spacings.
 Minimal mode
trapping could be an indicator that sdB stars are not as chemically
stratified as models usually presume. \citet{hu09} examined the effects
of diffusion on period spacing and their Fig.~4 shows damped mode
trapping, though for $\ell =3$ modes. Further diffusion may work
to remove a sharp mode-trapping boundary. Another possibility could
be thermohaline convection, caused
by an inverse $\mu$-gradient, as described by \citet{th09} which could increase
mixing and reduce chemical stratification. 

Our assignments as $\ell =1$ or 2 modes also adds constraints to
driving models. While Paper~IV produced models with
driven modes in the correct period range, 
previous modeling work \citep[including ][ among others]
{hu09,jeff07,font} had difficulties. Those models preferentially
found $\ell\geq 4$ to be driven (also at temperatures cooler than
observed). This is contradicted by our results,
which clearly follow Eqns.~3 and 4, indicating $\ell =1$ and 2 modes. 

Prior to space-based data such as \emph{Kepler} and CoRoT, it seemed unlikely
that sdB
asteroseismology using $g-$modes to probe the core would bear fruit.
The discovery of equal period spacings will now have changed that as we
can readily correlate modes with periodicities. The forward
method of mode assignment
is no longer necessary for these stars, which now provide a new 
modeling challenge. That challenge will be to model 
stars like KIC10670103, KIC10001893, and KIC10553698 which
have lengthy sequences of successive overtones,
equal period spacings which show minimal indications of mode trapping,
and provide tens of periods with secure mode assignments each.

We anticipate that once longer-duration \emph{Kepler} data are
available, many more pulsation periods will be detected. Already
there are typically too many periods to be accounted for solely
using $\ell =1$ and 2 modes and that problem will be compounded.
It is anticipated that the extra periodicities will be accounted
for using higher degree modes. Such an event will require more
sophisticated techniques and tests for assigning modes to
periodicities. However, the relatively simple tests of this paper
have been sufficient to confirm that regular period spacings in $g-$mode
sdB pulsators exist and provide useful constraints which stellar
models can now aspire to fit.

ACKNOWLEDGMENTS: Funding for this Discovery mission is provided by 
NASA's Science Mission
Directorate. The authors gratefully acknowledge the entire \emph{Kepler}
team, whose efforts have made these results possible. 
MDR was partially funded by a Missouri State University Faculty Research
Grant. ACQ is supported by
the Missouri Space Grant Consortium, funded by NASA. 
The research leading to these results has received funding from the
European Research Council under the European Community's Seventh Framework
Programme (FP7/2007--2013)/ERC grant agreement n$^\circ$227224 (PROSPERITY) and
 from the Research Council of K.U.Leuven (GOA/2008/04).
AB gratefully acknowledges
support from the Polish Ministry under grant No. 554/MOB/2009/0.

\section{On-line Material}
\setcounter{table}{3}
\begin{table}
\caption{Periods and period spacings for KIC2697388.
Identifications (column 1) and periods are those from Paper~III (Reed
et al. 2010).
Columns 3, 4, and 5 provide the mode degree $\ell$ and the overtone fit
to $P_{\ell}=P_{\ell o}+n\cdot\Delta 
P_{\ell}$ where n is arbitrarily chosen such that there are no negative values.
Column 6 provides the difference between the observed and asymptotic
relation period, column 7 lists the fractional period differences and
column 8 is the observed spacing $P(n_{\ell ,i})-P(n_{\ell ,j})/(i-j)$ 
It is ambiguous whether $\ell =1$ or 2 modes should be associated with f25, 
f23, and f11. f25 was not used for the $\ell =1$ fit as it is most likely
$\ell =2$.}
\label{tab04} 
\begin{tabular}{lccccccc}
\hline
ID & Period  & $\ell$ & n$_1$ & n$_2$ & $\delta P$ & $\delta P/\Delta P$ & Spacing  \\
 & (sec) &  &  & & (sec) &  &  (sec)\\ \hline
f36& 1124.964 & 1 & 0  &   & -0.373 &-0.002 & - \\
f35& 1279.784 &        - & - & - & - & - & - \\
f34& 1349.752 & 1 & 1  &   & -15.680& -0.065 & 224.8 \\
f34& 1349.752 & 2 &    & 5 &-2.578  &0.019 & - \\
f33& 1958.219 & - & -  & - & - & -  & - \\
f32& 1960.990 & - & -  & - & - & -  & - \\
f31& 1963.985 & - & -  & - & - & -  & - \\
f30& 2757.118 & 2 & -  & 15& 19.393 &  0.140 & 140.7 \\
f29& 3008.732 & 2 & -  & 17&-6.072  & 0.044 & 125.8 \\
f28& 3517.219 & 1 & 10 & - & -8.790 & -0.037 & 254.2 \\
f27& 3700.224 & 2 & -  & 22& -7.277 & -0.053 & 138.3 \\
f26& 3757.261 & 1 & 11 & - & -8.815 & -0.037 & 240.0 \\
f25& 3980.103 & 1 & 12 & - &-26.040 &-0.108  & 273.0 \\
f25& 3980.103 & 2 & -  & 24&-4.477  & 0.032 & 139.9 \\
f24& 4125.623 & 2 & -  & 25& 2.503  & 0.018 & 145.5 \\
f23& 4253.082 & 1 & 13 & - & 6.872  &  0.029 & 247.9 \\
f23& 4253.082 & 2 & -  & 26& -8.577 &  -0.062 & 127.5 \\
f22& 4536.116 & 2 & -  & 28&-2.622  &-0.019 & 141.5\\
f21& 4750.550 & 1 & 15 & - & 24.206 & 0.101 & 248.7 \\
f20& 4983.073 & 1 & 16 & - & 16.661 & 0.069 & 232.5 \\
f19& 5374.965 & 2 & -  & 34& 4.990  & 0.036 & 139.8 \\
f18& 5506.391 & 2 & -  & 35&-2.123  &-0.015 & 131.4 \\
f17& 5679.474 & 1 & 19 & - &-7.139  & -0.030 & 232.1 \\
f16& 5922.022 & 1 & 20 & - &-4.658  &-0.019 & 242.5 \\
f15& 6393.989 & - & -  &  - & - & - & - \\
f14& 6398.864 & 1 & 22 & - &-7.951  &-0.033 & 238.4 \\
f13& 6627.321 & 2 & -  & 43& 10.491 & 0.076 & 140.1 \\
f12& 6637.021 & 1 & 23 & - & -9.861 & -0.041 & 238.2 \\
f11& 6895.323 & 1 & 24 & - &  8.404 & 0.035  & 258.3 \\
f11& 6895.323 & 2 & -  & 45& 1.414  & 0.010 & 134.0 \\
f10& 7043.651 & 2 & -  & 46& 11.202 & 0.081 & 148.3 \\
f9 & 7294.763 & 2 & -  & 48& -14.764&-0.107 & 125.6 \\
f8 & 7607.932 & 1 & 27 & - & 0.781  & 0.003 & 237.5 \\
f7 & 7791.189 & - & -  &  - & - & - & - \\
f6 & 8312.182 & 1 & 30 & - & -15.170&-0.063 & 234.8 \\
f5 & 8367.893 & - & -  &  - & - & - & - \\
f4 & 8799.008 & 1 & 32 & - & -8.479 &-0.035 & 243.4 \\
f3 & 8970.500 & 2 & -  & 60&-1.501  &-0.011 & 139.6 \\
f2 & 10248.154 & 1& 38 & - & 0.264  &0.001 & 241.5 \\
f1 & 11222.207 & 1& 42 & - & 14.048 &0.059 & 243.5 \\ \hline
\end{tabular}
\end{table}

\begin{table}
\caption{Same as Table~4 for KIC2991403.
Identifications (column 1) and periods are those from Paper~V (Kawaler
et al., 2010).}
\begin{tabular}{lccccccc}
\hline
ID & Period  & $\ell$ & n$_1$ & n$_2$ &$\delta P$ & $\delta P/\Delta P$ & Spacing  \\
 & (sec) & &  & & (sec) &  &  (sec)\\ \hline
f16 & 2709.9  & 1 & 0 & -& 1.328 &0.005& - \\
f15 & 2981.4  & 1 & 1 & -& 4.304 &0.016& 271.5 \\
f14 & 2986.6  & 2 & - & 9 & 10.959& 0.071 & - \\
f13 & 2991.8   & - & - & - &    -   &     -   & -     \\
f12 & 3233.3   & - & - & - &    -   &     -   & -     \\
f11 & 3244.8  & 1 & 2 & - & 0.820& 0.003  & 251.9 \\
f10 & 3374.7  & - & - & - &    -  &  -      & -     \\
f9 & 3504.9  & 1 & 3 & -  & -9.244& -0.034& 260.1 \\
f8 & 3519.1  & - & - & - &    -   &    -    & -     \\
f7 & 3693.5   & - & - & - &    -   &     -   & -     \\
f6 & 3781.5  & 1 & 4 & -  & -1.167& -0.004  & 276.7 \\
f5 & 4326.6  & 1 & 6 &  - & 6.885 & 0.026  & 272.6 \\
f4 & 4337.3  & 2 & - & 18 & -22.908 & -0.149 &150.1\\
f3 & 5124.0  & 1 & 9 & -  & -1.286 & -0.005 & 265.8 \\
f2 & 5136.5  & 2 & - & 23 &  7.088 &  0.046 & 159.8\\
f1 & 6365.0  & 2 & - & 31 &  4.861 & 0.032  & 153.6 \\ \hline
\end{tabular}
\end{table}

\begin{table}
\caption{Same as Table~4 for KIC3527751.
Identifications (column 1) and periods are those from Paper~III (Reed et al.
2010).
Periods f25, f23, f5, f4, f3, and f2 are listed twice as they could be associated
with either $\ell =1$ or 2 modes. We suggest that f23 is $\ell =1$
and that f5, f4, f3, and f2 are all
$\ell =2$ modes and they were not used in the $\ell =1$ fit.
\label{tab04}}
\begin{tabular}{lccccccc}
\hline
ID & Period  & $\ell$ & n$_1$ & n$_2$ & $\delta P$ & $\delta P/\Delta P$ & Spacing  \\
 & (sec)& &  & & (sec) &  &  (sec)\\ \hline
f38& 984.427   & 1 & 0 & -& 7.425 & 0.028 & - \\
f37& 997.536  &-&-&-&-&-&-\\
f36& 1072.455  & -&-&-&-&-&-\\
f35& 1118.081  &-&-&-&-&-&-\\
f34& 1290.822  &-&-&-&-&-&-\\
f33& 1334.717  & 2 &- & 5 & -8.401 & -0.055 & - \\
f32& 1344.234  &-&-&-&-&-&-\\
f31& 1344.720  &-&-&-&-&-&-\\
f30& 1389.787  &-&-&-&-&-&-\\
f29& 1506.672  & 1 & 2 & -& -2.538 & -0.010 & 261.123 \\
f28& 1765.523  & 1 & 3 & -& -9.790  & -0.037 & 258.851 \\
f27& 1838.307  &-&-&-&-&-&-\\
f26& 2264.793  & 2 &- & 11 & 0.243 & 0.002 & 155.013 \\
f25& 2571.681  & 1 & 5 & -& -1.943 & -0.007 & 268.719 \\
f25& 2571.681  & 2 &- & 13 &-0.012 & 0.001 & 153.444 \\
f24& 2728.982  & 2 &- & 14 & 3.717 & 0.024 & 157.301 \\
f23& 3633.516  &  1 & 10 & -& -4.523 & -0.017 & 265.459 \\
f23& 3633.516  & 2 &- & 20 & -13.181 & -0.086 & 150.756 \\
f22& 3799.665  & 2 &- & 21 &-0.604 & 0.004 &  166.149\\
f21& 3811.326  &-&-&-&-&-&-\\
f20& 3910.932  & 1 & 11 & -& 6.789 & 0.026 & 277.416 \\
f19& 3951.085  & 2 &- & 22 & -2.756 & -0.018 & 151.420\\
f18& 4263.556  & 2 &- & 24 & 2.571 & 0.017 & 156.236 \\
f17& 4574.895  & 2 &- & 26 & 6.767 & 0.044 & 155.670 \\
f16& 4588.472  &-&-&-&-&-&-\\
f15& 4702.076  & 1 & 14 & -& -0.378 & -0.001 & 263.715 \\
f14& 4719.823  & 2 &- & 27 & -1.877 & -0.012 & 144.928 \\
f13& 4875.586  & 2 &- & 28 & 0.314 & 0.002 & 155.763 \\
f12& 5523.277  & 1 & 17 & -& 22.512 & 0.085 & 273.734 \\
f11& 5816.402  & 2 &- & 34 & 19.698 & 0.128 & 156.803 \\
f10& 5956.947  & 2 &- & 35 & 6.671 & 0.043 &  140.545\\
f9 & 6254.758  & 2 &- & 37 & -2.662 & -0.017 & 148.906 \\
f8 & 6285.436  & 1 & 20 & -& -13.640 & -0.051 & 254.053 \\
f7 & 7091.397  & 1 & 23 & -& -5.990  & -0.023 & 268.654 \\
f6 & 7365.565  & 1 & 24 & -&  2.075 &  0.008 & 274.168 \\
f5 & 7932.131  & 1 & 26 & -& 36.433 & 0.137 & 283.283 \\
f5 & 7932.131  & 2 &- & 48 & -14.580 & -0.095 &  152.488\\
f4 & 8415.768  & 1 & 28 & -& -12.137 & -0.046 & 241.819 \\
f4 & 8415.768  & 2 &- & 51 & 8.341 & 0.054  & 161.212 \\
f3 & 8724.069  & 1 & 29 & -& 30.060 & 0.113 & 308.301 \\
f3 & 8724.069  & 2 &- & 53 & 9.499 & 0.062 & 154.151 \\
f2 & 9480.648  & 1 & 32 & -& -11.672 & -0.044 & 252.193 \\
f2 & 9480.648  & 2 &- & 58 & -1.782 & -0.012 & 151.316 \\
f1 &10852.611  & 2 &- & 67 & -11.966 & -0.078 &  152.440\\ \hline
\end{tabular}
\end{table}

\begin{table}
\caption{Same as Table~4 for KIC5807616 (KPD1943).
Identifications (column 1) and periods are those from Paper~III (Reed et al.
2010).
We include the \emph{suggested} frequencies of Paper~III  (labeled with an
s in column 1).
Periods f14, f10, f9, and f6 are listed twice as they could be associated
with either $\ell =1$ or 2 modes. We suggest that f14 and f9 are $\ell =2$.
The last two columns provide the associated $\ell$ and $n$ values from 
(van Grootel et al. 2010).
\label{tab05}}
\begin{tabular}{lcccccccccc}
\hline
ID & Period  & $\ell$ & n$_1$ & n$_2$ & $\delta P$ & $\delta P/\Delta P$ & Spacing & VVG ID \\
 & (sec) &  & & & (sec) &  &  (sec) & $\ell$ & $n$ \\ \hline
s25&2346.420 &  -  & - & - & - & - & - &  2 &15\\
f18&2475.513 &  2  & - &16 &  2.381 &  0.017 &- & 2&16 \\
f17&2540.590 &  -  & - & - & - & - & - & 1 & 9\\
s24&2765.289 &  -  & - & - & - & - & - & 4 & 34\\
f16&2774.381 &  -  & - & - & - & - & - & 2 &18\\
f15&3162.719 &  2  & - &21& -6.063 & -0.044 &137.44  & 1 & 11\\
f14&3314.876 &  1 & 12& - & 16.512 & 0.068& - &2&22 \\
f14&3314.876 &  2 & - & 22&  6.964 &  0.050 & 152.16 & 2 &22 \\
s23&3446.011 &  2 & - & 23& -1.031 & -0.008 & 131.14 & 2 &23\\
f13&3786.345 &  1 & 14& - &  43.744&  0.015 & 235.73   &  1 &14 \\
f12&4002.153 &  2 & - & 27& -1.050 & -0.008  & 139.13 & 1 & 15\\
f11&4027.352 &  1 &15& - &   2.632 &  0.011 & 241.01 &  2 & 27\\
f10&4272.329 &  1 &16& - &   5.490 &  0.023 & 244.98 & 2 & 29\\
f10&4272.329 &  2 & - & 29& -9.494 & -0.068  & 134.91 & 2 &29\\
s22&4426.714 &  2 & - & 30&  5.761 &  0.041 & 141.40 & 2 &30\\
f9 &4963.086 &  1 &19 & - &-30.109 &-0.124  & 230.25  &2 &34\\
f9 &4963.086 &  2 & - & 34 &-14.387& -0.103 & 134.09 & 2 &34\\
f8 &5012.561 &  - & - & - & - & - & - &   1 &19\\
f7 &5726.085 &  1 &22 & - &  6.534  &  0.027 & 254.33  &1 & 22\\
f6 &5958.808 &  1 &23 & - & -2.862  &-0.012&  232.72 &  2&41 \\
f6 &5958.808 &  2 & - & 41 & 7.424  & 0.053 & 142.25 & 2 &41\\
f5 &6198.841 &  1 &24 & - & -4.948 &-0.020  &  240.03   & 1 & 24\\
f4 &6678.527 &  1 &26 & - & -9.499 &-0.039  &  239.84    & 2 &46\\
f3 &7055.779 &  - & - & - & - & - & - &     2 & 49\\
f2 &8372.266 &  1 &33 & - &-10.591 &-0.044  &  241.96    &2 & 58\\
f1 &9132.311 &  1 &36 & - & 23.098 &  0.095 & 253.35    &1 & 36\\ \hline
\end{tabular}
\end{table}

\begin{table}
\caption{Same as Table~4 for KIC7664467
Identifications (column 1) and periods are those from Paper~III (Reed et al.
2010).
 We include a \emph{suggested} frequency of Paper~III (labeled with an
s in column 1).
\label{tab06}}
\begin{tabular}{lccccccc}
\hline
ID & Period  & $\ell$ & n$_1$ & n$_2$ & $\delta P$ & $\delta P/\Delta P$ & Spacing  \\
 & (sec) &  & & & (sec) &  &  (sec)\\ \hline
f1 &4050.680 & 2& - &11 & 14.96 & 0.10 & - \\
f2 &4135.197 & 1& 0 & - &11.212 & 0.043& - \\
s7 &4640.562 & 1& 2 & - & -3.464& -0.013& 252.68\\
f3 &5156.458 & 1& 4 & - & -7.609& -0.029& 257.95\\
f4 &5688.625 & 1& 6 & - &  4.517&  0.017& 266.08\\
f5 &7487.792 & 1& 13& - & -16.459&-0.063 & 257.02\\
f6 &9076.175 & 1& 19& - & 11.801& 0.045& 264.73\\ \hline
\end{tabular}
\end{table}

\begin{table}
\caption{Same as Table~4 for KIC7668647
Identifications (column 1) and periods are those from Paper~VII (Baran et al.
2011).
s21 is not listed in Paper~VII, but is suggested by us. Periods f13 and
f5 are listed twice as they could either $\ell =1$ or 2 modes.
\label{tab07}}
\begin{tabular}{lccccccc}
\hline
ID & Period  & $\ell$ & n$_1$ & n$_2$ & $\delta P$ & $\delta P/\Delta P$ & Spacing  \\
 & (sec) &  & & & (sec) &  &  (sec)\\ \hline
f18& 2891.88  &  1 &0  & - & -11.11  & -0.045 & - \\
f17& 3387.730 &  2 & - & 12&  1.76   & 0.012  & - \\
s21& 3392.21  &  1 &2  & - & -7.07   & -0.028 & 250.17\\
f16& 3820.65  &  2 & - & 15&  0.56   &  0.004 & 144.31\\
f15& 4112.56  &  2 & - & 17&  3.06   &  0.021 & 145.96\\
f14& 4146.91  &  1 &5  & - & 3.19    & 0.013  & 251.57\\
f13& 4414.39  &  1 &6  & - & 22.52   & 0.091  & 267.48\\
f13& 4414.39  &  2 & - &19 &  15.474 & 0.107  & 150.92\\
f12& 4889.65  &  1 &8  & - & 1.48    & 0.006  & 237.63\\
f11& 4967.24  &  - & - & - & -       & -      & -  \\
f10& 5099.84  &  2 & - & 24& -22.61  & -0.156 & 132.60\\
f9 & 5140.81  &  1 &9  & - & 4.49    & 0.018  & 251.16\\
f8 & 5288.87  & -  & - & - & -       & -      &  - \\
f7 & 5877.88  &  1 &12 & - &  -2.89  & -0.012 & 245.69\\
f6 & 6288.29  &  2 & - & 32&    8.179 & 0.057 & 148.56\\
f5 & 6863.01  &  1 &16 & - &  -10.34  & -0.042 & 246.29\\
f5 & 6863.01  &  2 & - & 36&    4.08  & 0.028 & 143.68\\
f4 & 7605.87  &  1 &19 & - &  -11.92  & -0.048 & 247.62\\
f3 & 8367.25  &  1 &22 & - &  5.01    & 0.020 & 253.79\\
f2 & 8629.77  &  1 &23 & - &  19.386  & 0.078 & 262.52\\
f1 & 9093.93  &  1 &25 & - &  -12.75  & -0.051& 232.08\\ \hline
\end{tabular}
\end{table}

\begin{table}
\caption{Same as Table~4 for KIC8302197.
Identifications (column 1) and periods are those from Paper~VII (Baran et al.
2011).
Suggested periods (labeled with an s in column 1) are not listed in Paper~VII.
\label{tab08}}
\begin{tabular}{lcccccc}
\hline
ID & Period  & $\ell$ & n$_1$ & $\delta P$ & $\delta P/\Delta P$ & Spacing  \\
 & (sec) &  & & (sec) &  &  (sec)\\ \hline
f7  &   3269.75  &1&    0 & -8.17 & -0.032 & -\\
f6  &   3530.54 &1&    1 &  -5.09 & -0.020 &  260.78\\
s9  &   4577.5  &1&    5 &        11.07 & 0.043  &  261.74 \\
f5  &   5347.92 &1&    8 &   8.39 & 0.033  & 256.81\\
f4  &   5607.43 &1&    9 &  10.19 & 0.040  & 259.50 \\
f3  &   6109.39 &1&    11&  -3.25 & -0.013 &  250.98\\
f2  &   6868.71 &1&    14&  -17.02& -0.066 &  253.11\\
s8  &   7403.6   &1&    16&        2.47 & 0.010 & 267.45  \\
f1  &   7917.94 &1&    18 &  1.41 &  0.005&  257.17\\\hline
\end{tabular}
\end{table}

\begin{table}
\caption{Same as Table~4 for KIC9472174 (2M1938+4603).
Identifications (column 1) and periods are those from (\O stensen et al.
2010). They
grouped periods into regions labeled by A, B, and so on. We only
included those in region A, as those are the $g-$mode periods. fA20 is
also listed as $\ell =2$ though $\ell =1$ is a better fit.
\label{tab09}}
\begin{tabular}{lccccccc}
\hline
ID & Period  & $\ell$ & n$_1$ & n$_2$ & $\delta P$ & $\delta P/\Delta P$ & Spacing  \\
 & (sec) &  & & & (sec) &  &  (sec)\\ \hline
fA20 & 2158.441 & 1 & 0 &- & -8.469 & -0.033 & - \\
fA20 & 2158.441 & 2 &- & 6 & 18.184 & 0.123 & - \\
fA19 & 2303.443 & 2 &- & 7 & 15.490 & 0.105 & 145.00  \\
fA18 & 2692.512 & 1 & 2 &- & 14.348 & 0.056 & 267.04  \\
fA17 &   2824.900 &- &- &- &- &- &- \\
fA16 & 2951.098 & 1 & 3 &- & 17.307 & 0.068 & 258.59  \\
fA15 & 3134.262 & 2 &- & 13 & -39.867 & -0.270 & 138.47 \\
fA14 &  3418.37 &- &- &- &- &- &- \\
fA13 & 3456.036 & 1 & 5 &- & 10.991 & 0.043 & 252.47  \\
fA12 & 4444.656 & 1 & 9 &- & -22.897 & -0.090 & 247.16  \\
fA11 &  4567.474 &- &- &- &- &- &- \\
fA10 & 4634.736 & 2 &- & 23 & -16.353 & -0.111 & 150.05\\
fA9 & 5474.256 & 1 & 13 &- & -15.805 & -0.090 & 257.40\\
fA8 & 5853.966 & 2 &- & 31 & 21.309 & 0.144 & 152.40\\
fA7 & 6577.086 & 2 &- & 36 &  5.948 & 0.040 & 144.62\\
fA6 & 7452.603 & 2 &- & 42 &-4.711 & 0.032 & 145.92\\
fA5 & 8557.319 & 1 & 25 &- &-0.266 & 0.001 & 256.92\\
fA4 & 10208.828 & 2 &- & 61 & -54.710 &-0.370 & 145.06\\
fA3 &    10915.351 &- &- &- &- &- &- \\
fA2 & 17253.692 & 1& 59 &- &  4.790 & 0.019 & 255.78 \\
fA1 & 19884.934 & 2 &- & 121 & 21.155 & 0.143 & 148.86\\ \hline
\end{tabular}
\end{table}

\begin{table}
\caption{Same as Table~4 for KIC10001893.
Identifications (column 1) and periods are those from Paper~VII (Baran et
al. 2011).
Periods shortward of f14 can be associated with different modes using
slightly different period spacings. Both results are provided with Scheme 2
having a slightly better fit. Suggested periods s27, s28, and s29 
were detected using a rough fit to a wavelet analysis and require further work
to determine their validity. 
\label{tab10}}
\begin{tabular}{lccccccc|cccccc}
\hline
     &        & \multicolumn{6}{c}{Scheme 1} & \multicolumn{6}{c}{Scheme 2} \\
\hline
ID & Period  & $\ell$ & n$_1$ & n$_2$ & $\delta P$ & $\delta P/\Delta P$ & Spacing & $\ell$ & n$_1$ & n$_2$ & $\delta P$ & $\delta P/\Delta P$ & Spacing \\
 & (sec) &  & & & (sec) &  &  (sec)&  & & & (sec) &  &  (sec)\\ \hline
f26& 1023.05&1& 0&-&13.35 & 0.050   & -       & 2&-&3 & 0.24    &0.001 & - \\
f25& 1472.43&2&-&6 & -26.14& -0.169 & -       & 1& 2& -& -19.57 &-0.127 & - \\
f24& 2555.15&2&-&13& -24.65& -0.160 & 154.67  & 1& 6& -& -10.99 &-0.071 & 270.68 \\
f23& 2780.21&2&-& 14& 45.96& 0.298  & 225.07  & -& -& -&  -   & -& -\\
f22& 3086.54&2&-& 16& 43.35& 0.281  & 153.16  & 1& 8& -& -16.66 & -0.108 & 265.70 \\
f21& 3348.33&2&-& 18& -3.77& -0.024 & 130.90  & 2& -&18& 4.36 & 0.028 & 155.02 \\
f20& 3496.20&2&-& 19&-10.36& -0.067 & 147.87  & 2& -&19& -2.51& -0.016& 147.87 \\
f19& 3645.63&1&10&-& -24.12& 0.091  &  262.26 & 1&10& -& 5.373& 0.020 & 279.55 \\
f18& 3802.95&2&-& 21&-12.54& -0.081 & 153.38  & 2& -&21& -5.25& -0.016& 153.38 \\
f17& 3919.22&1&11&-& -16.53& 0.062  & 273.59  & 1&11& -& 10.43 & 0.039 & 273.59 \\
f16& 4746.35&1&14&-&  12.59&  0.047 & 275.71  & 2& -&27& 9.87 & 0.036 & 130.96 \\
f15& 4884.89&2&-&28 &-11.84 &-0.077   &154.56 & 2& -&28& -6.53 & -0.042 & 154.56 \\
f14& 5022.31&1&15&-&  22.54&  0.085 & 275.96  & 1&15& -& 39.39 & 0.148 & 275.77 \\
s29& 5038   &2&-&29& -13.18& -0.085 & 153.12  & 2&- & 29& -8.15 & -0.053 & 153.12 \\
f13& 5288.98&1&16&-&  23.20&  0.087 & 266.66  & 1&16& -&37.52 & 0.141 & 266.66 \\
f12& 5540.21&1&17&-&   8.43&  0.032 & 251.23  & 1&17& -& 20.21 & 0.076 & 251.23 \\
f11& 5792.81&1&18&-& -4.98&-0.019   & 252.60  & 1&18& -& 4.28 & 0.016 & 252.60 \\
s28& 5803   &2&- &34&-20.49&-0.133  & 154.00  & 2&-& 34 & -16.9 & -0.109 & 153.0 \\
s27& 5956   &2& -&35&-21.96&-0.142  & 153.00  & 2&-& 35& -18.62 & -0.121 & 153.0 \\
f10& 6049.56&1&19&-& -14.22&-0.053  & 256.76  & 1&19& -&  -7.49 & -0.028 & 256.76 \\
f9 & 6329.06&1&20&-& -0.73&-0.003   & 279.50  & 1&20& -&  3.48 & 0.013 & 279.50 \\
f8 & 6576.60&1&21&-& -19.19&-0.072  & 247.54  & 1&21& -& -17.52 & -0.66 & 247.54 \\
f7 & 6845.23&1&22&-& -16.57&-0.062  & 268.63  & 1&22& -& -17.42 & -0.065 & 268.63 \\
f6 & 7112.78&1&23&-& -15.02&-0.056  & 267.14  & 1&23& -& -18.40 & -0.069 & 267.55 \\
f5 & 7388.93&1&24&-& -4.897 &-0.018 & 276.14  & 1&24& -& -10.80 & -0.041 & 276.14 \\
f4 & 7659.85&1&25&-&   0.03&  0.001 & 270.92  & 1&25& -& -8.41 & -0.032& 270.92 \\
f3 & 7942.18&1&26&-&   16.37&  0.062& 282.34  & 1&26& -& 5.40 & 0.020 & 282.34 \\
f2 & 8743.58&1&29&-&   19.74&  0.074& 267.13  & 1&29& -& 1.19 & 0.004 & 267.13 \\
f1 & 12899.51&-& - & - & - & - & -& -& -& -& - & - & -\\ \hline
\end{tabular}
\end{table}

\begin{table}
\caption{Same as Table~4 for KIC10553698.
Identifications (column 1) and periods are those from Paper~VII (Baran
et al. 2011).
Periods f36, f30, f20, f15, and f8 could be matched by either $\ell =1$
or 2, though f30 and f9 are most likely $\ell =1$ and f9 was not used
in the $\ell =2$ fit.
\label{tab11}}
\begin{tabular}{lccccccc}
\hline
ID & Period  & $\ell$ & n$_1$ & n$_2$ & $\delta P$ & $\delta P/\Delta P$ & Spacing  \\
 & (sec) &  & & & (sec) &  &  (sec)\\ \hline
f36& 2238.30  & 1 & 0 &- & 8.17    &0.030 & -  \\
f36& 2238.30  & 2 & -& 6 &  6.78   & 0.043 & -  \\
f35& 2382.81  & 2 & -& 7 & 1.74     & 0.011  & 144.50   \\
f34& 2518.16  & 1 & 1 &- & 16.88   & 0.062 & 279.86  \\
f33& 2545.22  & 2 & -& 8 &-5.40    &-0.034 & 162.41  \\
f32& 2759.15  & 1 & 2 &- & -13.28  & -0.049 & 240.99   \\
f31& 2867.13  & 2 & - & 10 & 12.35 & 0.079 & 160.96 \\
f30& 3036.87  & 1 & 3 &- & -6.715   & -0.025 & 277.72   \\
f30& 3036.87  & 2 & -& 11 & 21.95   & 0.140 & 169.74  \\
f29& 3124.89  & -& -& -& -& - &- \\
f28& 3294.19  & 2 & - & 13 & -30.17 & -0.193 & 142.35 \\
f27& 3308.79  & 1 & 4 &- & -5.966   & -0.025 & 271.91   \\
f26&  3447.91 & - & -& -& -& - &- \\
f25 & 3482.53 & 2 & -& 14 & -2.43    & -0.015 & 188.35   \\
f24 & 3587.68 & 1 & 5 & - & 1.78 & 0.007 & 278.89 \\
f23 & 3611.47 & 2 & -& 15 & -33.17   & -0.193 & 128.916  \\
f22 & 3859.72 & 1 & 6 &- & 2.68    & 0.010 & 272.05   \\
f21 & 4058.55 &- & -& -& -& - &- \\
f20 & 4123.14 & 1 & 7 &- & -5.05    & 0.019 & 263.42   \\
f20 & 4123.14 & 2 & -& 18 &11.47    & 0.073 & 170.56   \\
f19 & 4392.03 & 1 & 8 &- & -7.31    & -0.027 & 268.89  \\
f18 & 4673.52 & 1 & 9 &- & 3.03     &  0.011 & 281.49   \\
f17 & 4806.31 & -& -& -& -& - &- \\
f16 & 4951.42 & 1 & 10 &- & 9.77    & 0.036 & 277.89    \\
f15 & 5202.87 & 1 & 11 &- & -9.93   & -0.037 & 251.45   \\
f15 & 5202.87 & 2 & -& 25 & -5.57   & -0.036 & 159.14   \\
f14 & 5384.23 & 2 & -& 26 & 19.12   & 0.122 & 161.162  \\
f13 & 5703.13 & 2 & -& 28 & 24.66   & 0.157 & 159.45  \\
f12 & 5955.79 & -& -& -& -& - &- \\
f11 & 6223.85 & -& -& -& -& - &- \\
f10 & 6473.08 & 2&-& 33 & 11.21 & 0.072 & 153.99 \\
f9  & 6742.50 & 2 & - & 35 & -32.72 & -0.209 & 154.08 \\
f8  & 7116.77 & 1 & 18 &- &  5.93    &  0.022 & 273.42   \\
f8  & 7116.77 & 2 & -& 37 & 28.19   & 0.180 & 160.92   \\
f7  & 7269.10 & 2 & -& 38 & 23.84   & 0.152 & 156.60   \\
f6  & 7765.49 & -& -& -& -& - &- \\
f5  & 8330.75 & 2 & - & 45 & -11.28 & -0.072 & 151.66 \\
f4  & 8591.40 & -& -& -& -& - &- \\
f3  & 9115.72 & 2 & -& 50 & -9.70   & -0.062 & 157.00   \\
f2  & 9384.20 & -& -& -&- & - &- \\
f1  & 9588.52 & 2 & -& 53 & -6.95   & -0.044 & 157.60   \\ \hline
\end{tabular}
\end{table}

\begin{table}
\caption{Same as Table~4 for KIC010670103.
Identifications (column 1) and periods are those from Paper~III (Reed
et al. 2010).
\label{tab12}}
\begin{tabular}{lccccccc}
\hline
ID & Period  & $\ell$ & n$_1$ & n$_2$ & $\delta P$ & $\delta P/\Delta P$ & Spacing  \\
 & (sec) &  & & & (sec) &  &  (sec)\\ \hline
f28 &  4920.1  &  2 & - & 8  & 3.6   & 0.024 &   -   \\
f27 &  5061.6  &  2 & - & 9  & -0.5  & -0.004& 141.5 \\
f26 &  5208.5  &  2 & - & 10 & 0.7   & 0.005 & 146.9 \\
f25 &  5353.1  &  2 & - & 11 & -0.2  & -0.002& 144.6 \\
f24 &  5493.3  &  2 & - & 12 & -5.6  & -0.039& 140.2 \\
f23 &  6081.9  &  2 & - & 16 & 0.6   & 0.004 & 147.2 \\
f22 &  6484.6  &  1 & 0 & -  & -17.8 & -0.071&   -   \\
f21 &  6996.0  &  1 & 2 & -  & -8.6  & -0.034& 255.7 \\
f20 &  7106.8  &  2 & - & 23 & 6.3   & 0.044 & 146.4 \\
f19 &  7241.2  &  2 & - & 24 & -4.9  & -0.033& 134.4 \\
f18 &  7515.8  &  1 & 4 & -  & 8.9   & 0.035 & 259.9 \\
f17 &  7761.7  &  1 & 5 & -  & 3.7   & 0.015 & 245.9 \\
f16 &  8259.5  &  1 & 7 & -  & -0.8  & -0.003& 248.9 \\
f15 &  8758.8  &  1 & 9 & -  & -3.8  & -0.015& 249.7 \\
f14 &  9012.6  &  1 & 10 & - & -1.1  & -0.004& 253.8 \\
f13 &  9265.8  &  1 & 11 & - & 1.0   & 0.004 & 253.2 \\
f12 &  9506.1  &  1 & 12 & - & -9.8  & -0.039& 240.3 \\
f11 & 10271.0  &  1 & 15 & - & 1.7   & 0.007 & 255.0 \\
f10 & 10532.7  &  1 & 16 & - & 12.2  & 0.049 & 261.7 \\
f9 &  11316.3  &  1 & 19 & - & 42.44 & 0.169 & 261.2 \\
f8 &  11533.7  &  1 & 20 & - & 8.7   & 0.035 & 217.4 \\
f7 &  12278.2  &  1 & 23 & - & -0.2  & -0.001& 248.2 \\
f6 &  12788.2  &  1 & 25 & - & 7.6   & 0.030 & 255.0 \\
f5 &  13019.7  &  1 & 26 & - & -12.1 & -0.048& 231.5 \\
f4 &  13262.8  &  1 & 27 & - & -20.1 & -0.080& 243.1 \\
f3 &  13779.6  &  1 & 29 & - & -5.6  & -0.022& 258.4  \\
f2 &  15343.7  &  - & -  & - & -     & - & - \\
f1 &  16290.1  &  1 & 39 & - & -6.4 & -0.025& 251.1 \\
\hline
\end{tabular}
\end{table}

\begin{table}
\caption{Same as Table~4 for KIC11179657.
Identifications (column 1) and periods are those from Paper~V 
(Kawaler et al. 2010). Suggested frequencies are listed with an
s in column 1.
f1 is also listed as $\ell =2$ for completeness although it is most
likely $\ell =1$.
\label{tab13}}
\begin{tabular}{lccccccc}
\hline
\hline
ID & Period  & $\ell$ & n$_1$ & n$_2$ & $\delta P$ & $\delta P/\Delta P$ & Spacing \\
 & (sec) &  & & & (sec) &  &  (sec)\\ \hline
f11 & 2844.27  & 2 & - & 6 & 4.38 & 0.033 & - \\
f10 & 2956.26  & - & - & - &   -   &  -     & - \\
f9 & 2965.95  &  2 & - & 7 & -7.58 & -0.057 & 121.68 \\
f8 & 3240.38  & 2 & - & 9 &-0.44 &-0.003 & 137.215 \\
f7 & 3381.00  & 2 & - & 10 & 6.54 & 0.049 & 140.62 \\
f6 & 3503.48  & 2 & - & 11 & -5.03 & -0.038 & 122.48 \\
f5 & 3513.35  & 1 & 0 & - & 0.01 & 0.001 & - \\
f4 & 3522.81  & - & - &  -&  -   &  -    & - \\
s12 & 4314.70  & 2 & - & 17 & 4.72 & 0.035 & 135.2 \\
f3 & 5109.26  & 2 & - & 23 & -2.59 & -0.019 & 132.4 \\
f2 & 5130.40  & 1 & 7 & - &-0.09 &-0.001 & 231.0 \\
f1 & 5361.58  & 2 & - & 25 & -18.35 & -0.137 & 126.16 \\
f1 & 5361.58  & 1 & 8 & - & 0.07& 0.001 & 231.18 \\ \hline
\end{tabular}
\end{table}

\begin{table}
\caption{Same as Table~4 for KIC11558725.
Identifications (column 1) and periods are those from Paper~VII (Baran
et al. 2010).
Suggested frequencies s54 and s55 are not listed in Paper~VII, but
are suggested by us and listed with an s in column 1.} 
\label{tab14}
\begin{tabular}{lccccccc}
\hline
ID & Period  & $\ell$ & n$_1$ & n$_2$ & $\delta P$ & $\delta P/\Delta P$ & Spacing  \\
 & (sec) &  & & & (sec) &  &  (sec)\\ \hline
f44 &1403.78   &2 &- & 0 &    2.05  &0.014  & - \\
f43 &1406.85   &- & - & - & - & - & - \\
f42 &1519.33   &2 &- & 1 &  -24.97  &-0.175 & -115.55 \\
f41 &2558.20   &2 &- & 8 &  15.90   &0.112 & 148.41 \\
f40 &2613.64   &- & - & - & - & - & - \\
f39 &2741.74   &- & - & - & - & - & - \\
s55 &2834.64  &2 &- & 10 &  7.20   &0.051 & 138.22 \\
f38 &2855.62   &- & - & - & - & - & - \\
f37 &3100.84   &- & - & - & - & - & - \\
f36 &3116.80   &2 &- & 12 &   4.22  &0.030 & 139.65  \\
f35 &3211.96   &- & - & - & - & - & - \\
f34 &3250.93   &2 &- & 13 &  -4.22   &-0.030 & 134.13  \\
f33 &3276.54   &- & - & - & - & - & - \\
f32 &3377.28   &- & - & - & - & - & - \\
f31 &3388.92   &- & - & - & - & - & - \\
f30 &3530.80   &2 &- & 15 &  -9.49    &-0.067& 139.94  \\
f29 &3640.52   &1 & 5 &-&    -20.81  &-0.084 & - \\
f28 &3686.12   &2 &- & 16 &   3.26   & 0.023  & 155.32  \\
f27 & 3822.21  &2 & - & 17 & -3.12 & -0.022 & 136.20 \\
f26 &3969.80   &2&- &  18 &   1.80    &0.013 & 147.48  \\
f25 &4160.37   &1 & 7 &-&     5.51   & 0.022 & 260.66  \\
f24 &4359.19   &- & - & - & - & - & - \\
f23 &4405.37   &2 &- & 21 &  9.65   &  0.068 & 145.19 \\
f22 &4421.03   &1 & 8 &-&    19.40   & 0.079 & 260.66 \\
f21 &4663.89   &1 & 9 &-&    15.49  & 0.063 & 242.86 \\
f21 &4663.89   &2 &- & 23 &  -16.97   &-0.119 & 129.26 \\
f20 &4908.04   &1 & 10&-&    12.87   & 0.052 & 244.15  \\
f19 &5150.71   &1 & 11&-&    8.78    & 0.036 & 242.68  \\
f18 &5375.06   &1 & 12&-&   -13.64   &-0.055 &224.352 \\
f17 &5610.76   &1 & 13&-&   -24.72   &-0.100 &236.82  \\
f16 &5675.24   &- & - & - & - & - & - \\
f15 &5869.32   &1 & 14&-&   -12.92   &-0.052 &258.56 \\
f14 &6106.13 &- & - & - & - & - & - \\
f13 &6371.19   &1 &16 &-&   -4.60   &-0.019&265.05 \\
f13 &6371.19   &2 &- & 35 & -20.529  &-0.144  & 142.28 \\
f12 &6552.55   &2 &- & 36 & 18.27   & 0.128 & 143.15 \\
f11 &6626.11   &1 &17 &-&   3.57    & 0.014 & 254.93  \\
f10 &6892.92   &1 &18 &-&   23.60   & 0.096 & 266.80  \\
f9  &7121.59   &1 &19 &-&   5.52  & 0.022  &228.68 \\
f9  &7121.59   &2 &- & 40 & 17.03  & 0.119 & 150.08 \\
s54 &7354.49   &1 &20 &-&   -8.354 & -0.034 & 232.90  \\
f8  &7624.98   &1 &21 &-&   15.36  &  0.062 & 270.48  \\
f7  &8101.45   &1 &23 &-&   -1.70  &-0.007  &238.24  \\
f7  &8101.45   &2 &- & 47 & -1.11  &-0.008   & 140.81 \\
f6  &8612.79   &1 &25 &-&   16.10  & 0.065&255.67 \\
f5  &9567.17   &1 &29 &-&  -16.60  & -0.067&  238.60  \\
f4  &9900.75   &- & - & - & - & - & - \\
f3  &9990.93 & - & - & - & - & - &-\\
f2  &11094.93  &1 &35 &-&  42.168  & 0.171 &254.36 \\
f2  &11094.93  &2 &- & 68 & -1.61   &-0.011 & 142.55 \\
f1  &12798.82  &1 &42 &-&   21.04  &  0.085 &243.197 \\
f1  &12798.82  &2 &- & 80 & -8.58  &-0.060 & 141.99 \\ \hline
\end{tabular}
\end{table}

\begin{table}
\caption{Same as Table~4 for KPD0629-0016.
Identifications (column 1) and periods are those from (Charpinet
et al. 2010).
 Unlike Tables~4-16, these periods are ordered by descending amplitude.
Periods f8 and f12 are shown for both $\ell =1$ and 2.
\label{tab15}}
\begin{tabular}{lccccccc}
\hline
ID & Period  & $\ell$ & n$_1$ & n$_2$ & $\delta P$ & $\delta P/\Delta P$ & Spacing  \\
 & (sec) &  & & & (sec) &  &  (sec)\\ \hline
f7  & 2601.915 & 1 & 0 & -&-16.86  &-0.068 & - \\
f3  & 2838.557 & 1 & 1 & -&-27.38  &-0.111 & 236.64 \\
f5  & 3355.602 & 1 & 3 & -&-4.68  &-0.019 &  258.52 \\
f9  & 3513.892 & 2 & - & 14 & 1.89   &  0.013  & - \\
f2  & 3614.638 & 1 & 4  & -&  7.18  & 0.029 & 259.04 \\
f6  & 3662.012 & 2 & - &  15 & 7.26  & 0.051 & 148.12 \\
f8  & 4360.573 & 1 & 7 & -&  11.60  &  0.047 &  248.645 \\
f8  & 4360.573 & 2 & - &  20 & -7.88   & -0.055 & 139.71 \\
f4  & 4618.324 & 1 & 8 & -&  22.18  &  0.090 & 257.751 \\
f1  & 4871.190 & 1 & 9 & -&  27.89 &  0.113 &  252.866 \\
f10 & 4878.150 & - & - &  - & - & - & - \\
f15 & 6820.557 & 1 & 17 & -&-0.13  &-0.001  & 243.67    \\
f13 & 7301.076 & 1 & 19 & -&-13.95  &-0.056 & 240.26   \\
f11 & 7640.621 & 2 & - & 43 & -10.87  & -0.076 & 142.61 \\
f14 & 8304.588 & 1 & 23 & -&  0.88  &  0.004 & 250.88  \\
f16 & 8477.007 & - &  - & - & -& - & - \\
f12 & 8803.020 & 1 & 25 & -&  4.97&  0.020 & 249.22  \\
f12 & 8803.020 & 2 & - & 51 & 9.60 & 0.067 & 145.30\\
f17 & 10516.563 & 1 & 32 & -&-11.67  &-0.047 & 244.79  \\ \hline
\end{tabular}
\end{table}

\end{document}